\documentclass{article}
\usepackage{amsmath}
\usepackage{amssymb}
\usepackage[margin=1in]{geometry}
\usepackage[nosort]{cite}
\usepackage{graphicx}
\usepackage{color} 
\usepackage{bbm}

\usepackage[utf8]{inputenc}

\newcommand{\ba}{\begin{align}}
\newcommand{\ea}{\end{align}}

\newcommand{\nn}{\nonumber}
\newcommand{\lpangle}{(}
\newcommand{\rpangle}{)}

\begin{document}
	\begin{flushright}\footnotesize
		\texttt{
		}
	\end{flushright}
	\vspace{1cm}
	
	\begin{center}%
		{\LARGE{Classical solitons in anti-self-dual Yang-Mills theory
			}\par}%
		
		\vspace{15mm}

		\centerline{ 
			{Tristan McLoughlin}
		}
		
		\bigskip\bigskip
		
		\noindent{\em
		School of Mathematics \& Hamilton Mathematics Institute, \\
		Trinity College Dublin, Ireland\\
		[5pt]
	}
		\texttt{tristan@maths.tcd.ie}
		
		\par\vspace{14mm}
		
		\textbf{Abstract} \vspace{5mm}

		\begin{minipage}{14cm}
			\parindent=2em
			We study classical soliton solutions in split-signature anti-self-dual Yang-Mills theory. Using Zakharov's dressing method, we construct rational lump solitons, which are localised near two-dimensional surfaces, and wall solitons, which have codimension one. We further describe multi-phase solutions which have multiple walls meeting at trivalent junctions and form four-dimensional foam-like structures. Two-pole solutions describe soliton interactions and we find non-trivial colour-dressed phase shifts for wall interactions. We consider the four-dimensional analogue of Ward's lump scattering for the restricted kinematics of coincident poles and show that there are non-trivial interactions in the collinear limit.
			Finally, we show that the determinant of the soliton matrix provides a $\tau$-function which acts as a potential for gauge-invariant scalar quantities such as the Yang-Mills Lagrangian density.		
			
		\end{minipage}
		
	\end{center}

	\newpage
	\tableofcontents
	
	\section{Introduction}
	Anti-self-dual Yang-Mills (ASDYM) theory can be viewed as an ur-theory of integrable systems in the sense
	that many models descend from it via symmetry reduction \cite{Ward:1985gz} and its twistor formulation provides a framework for understanding structures that appear in otherwise disparate contexts \cite{Mason:1991rf}. 
	In addition, ASDYM has been of significant theoretical importance through its role in the study of scattering amplitudes since at least the work of Bardeen \cite{Bardeen:1995gk}, who suggested that the simplicity of certain perturbative gluon amplitudes is related to the integrable structures of ASDYM. 
	There has been renewed attention to this aspect of the theory due to its appearance in the celestial chiral algebra programme for computing amplitudes \cite{Costello:2021bah, Costello:2022wso} and the direct computation of non-trivial single-minus tree-level $n$-point amplitudes in half-collinear kinematics \cite{Guevara:2026qzd}. 

The construction of soliton solutions has generally been an important part of the study of integrable models, both for its own inherent interest and as a tool to probe the quantum theory. Notably for the sine-Gordon model, semi-classical quantisation of solitons \cite{Korepin:1975vd, Korepin:1975zu, Dashen:1975hd} and  classical time delays in soliton scattering \cite{Jackiw:1975im} played a role in determining the  exact factorised S-matrix \cite{Zamolodchikov:1978xm}. The understanding of solitons in higher-dimensional integrable systems is significantly less advanced although there has been substantial progress in a number of directions. One prominent example is the KP-II equation, which describes the evolution of two-dimensional small-amplitude waves in a weakly, negatively dispersive medium \cite{KP}. Its line-soliton solutions at a fixed time form complex web-like patterns. These solutions can be constructed using a Wronskian $\tau$-function, which can be expressed as a sum of exponentials, and can be mapped to a point in the real Grassmannian \cite{hirota2004direct, sato1981soliton}. A remarkable connection between the combinatorics of the totally positive and non-negative parts of real Grassmannians and regular KP-II solitons has been shown to exist and used to provide a classification of line-soliton solutions \cite{kodama2011kp, kodama2013deodhar, kodama2014kp}; see \cite{kodama2015full, betancourt2026combinatorial} for related work on the Toda lattice and Davey-Stewartson equation. 

A second direction concerns the three-dimensional integrable chiral model. Its version with a time-like preferred vector was first studied by Manakov and Zakharov \cite{Manakov:1981kz} and the space-like version by Ward \cite{Ward:1988ie}. Ward constructed multi-lump solutions and showed that for generic poles the solitons pass through each other without deflection or phase shift \cite{Ward:1988ie}. However, non-trivial scattering was numerically shown to occur for special kinematics \cite{Sutcliffe:1992dj} and Ward subsequently gave an analytical derivation in terms of double-pole solutions \cite{Ward:1995he}. In work which motivates our approach \cite{Leese:1989hj}, it was shown that additional non-trivial scattering occurs in the case of extended wave solitons  which interact non-trivially with both lumps and other extended waves. The factorisation properties of these three-dimensional soliton interactions have recently been studied by Cole \cite{Cole:2026iqj}.

ASDYM is also well known to have a rich space of classical solutions. In Euclidean space-time, instanton solutions  have long been studied \cite{Belavin:1975fg, Ward:1977ta, Atiyah:1977pw} and can be found via the ADHM construction \cite{Atiyah:1978ri}. Extended soliton solutions were constructed using 
the Darboux method \cite{Nimmo:2000aj} and shown to have an elegant formulation in terms of quasideterminants \cite{gel1991determinants, gel1992theory}; see also \cite{deVega:1987dj} for earlier work on extended plane-wave solutions. Subsequently, this was developed to identify Wronskian-type quasideterminant solutions whose action densities are localised on three-dimensional hyperplanes and it was shown that $n$-soliton solutions can be interpreted as intersecting walls with asymptotic phase shifts \cite{Gilson:2020neb, Hamanaka:2020pyc, Hamanaka:2021qsk, Huang:2022mqh, Hamanaka:2022ole, Hamanaka:2024ctw}. In closely related work, the Cauchy
matrix approach was applied to the ASDYM equations to construct Gramian-type solutions
\cite{Li:2021kxj, Li:2022fum} and it was proved that, for the case of split signature and unitary gauge group, the Wronskian and Gramian soliton solutions gave rise to equivalent gauge-invariant densities \cite{Li:2025dlc} in the one- and two-pole cases. The two formulations of the general solution were recently shown to be equivalent and to generate identical ASDYM potentials \cite{Huang:2026vvw, li2026equivalence}. Further recent work has developed a Sato-type description for the ASDYM hierarchy \cite{Li:2026tnj}.

In this work we will consider the computation of ASDYM soliton solutions in flat space-time with signature $(+--+)$. We will follow the approach of Zakharov and collaborators \cite{zakharov1978relativistically, zakharov1979integration}, often referred to as the Riemann-Hilbert method with zeros or the dressing method, in which solutions are generated by a matrix-valued function of the spectral parameter with specified poles and subject to a reality condition. This method was applied by Belavin and Zakharov to the construction of instanton solutions in the Euclidean theory \cite{Belavin:1978pa} and has also been extensively used in the three-dimensional integrable chiral model \cite{Ward:1990vc, Fokas:1998sp, Dai:2004fj}. We apply this method to construct the four-dimensional analogues of Ward's rational lumps, which live on two-surfaces in space-time, and Leese's exponential wave solutions, which live on three-surfaces, in SU$(2)$ ASDYM. This is a straightforward extension of the chiral model results and the extended waves are exactly those previously found by the Darboux method. 

However, one advantage of this approach is that it allows us to identify the determinant of the soliton matrix, which is central to the dressing method, as a $\tau$-function for this class of solutions. We mean this in the restricted sense that $\tau$ is a positive function whose logarithm acts as a potential for gauge-invariant densities that characterise solutions. In particular, we show that the Yang-Mills Lagrangian density, evaluated on this class of solutions with simple poles, can be computed as 
\begin{align}
	\mathcal{L}=-\Box^2 \log \tau~.
\end{align}
This result is strongly reminiscent of Osborn's formula for $\mathcal{L}$ as a double Laplacian of a scalar function constructed from ADHM data \cite{Osborn:1979bx, Osborn:1981yf}\footnote{For a detailed derivation see Appendix C of \cite{Dorey:2002ik}.}
which suggests that there may be a more general relation. We also find that for poles on the positive imaginary axis, and appropriate choices of the Wess-Zumino representative, the scalar $\sigma=\Box \log \tau$ is proportional to the four-dimensional WZW Lagrangian density \cite{Pohlmeyer:1979ya, Donaldson:1985zz, Nair:1990aa, Losev:1995cr}. This bears comparison to the relation between the KP field $u$ and $\tau$-function, $u(x,y,t)=2\partial_x^2 \log \tau$, and suggests that the $\tau$-function introduced in this work may play a role in developing a combinatorial description of ASDYM solutions\footnote{After this work was complete, the preprint \cite{li202621dim} appeared showing that the Yang-Mills Lagrangian density dimensionally reduced to 2+1 dimensions could be written in terms of a $\tau$-function given by the determinant of a Cauchy-type matrix and, for certain solutions, becomes a Gramian-type $\tau$-function of the KP equation.}. For our purposes here, its main use lies in providing a relatively transparent method for extracting physical properties from solutions. 

In the one-pole sector we construct multi-phase solutions describing rigid configurations of intersecting colour-dressed walls giving rise to a foam-like arrangement in four-dimensional space-time.  One novel feature, distinct from the three-dimensional chiral model,  is the presence of singular defects decorating our solutions at isolated points which, for an appropriate choice of regularisation, carry a finite action. In the two-pole sector, we are able to construct solutions which describe the non-trivial interactions of walls and compute a colour-dressed generalisation of the phase shift previously found in the three-dimensional model  \cite{Leese:1989hj} and in \cite{Hamanaka:2022ole, Li:2025dlc} for ASDYM. The colour dressing allows for configurations where the phase shift can vanish or change sign depending on the relative colour orientation of the walls. We additionally consider the scattering of lumps from walls and compute the resulting transverse displacement of the lump. Finally, in lump-lump scattering we find the same result as Ward with, for example, non-trivial two-lump scattering for head-on collisions \cite{Ward:1995he}.

\section{Anti-self-dual Yang-Mills theory}

 We start with non-Abelian gauge theory with gauge group SU$(N_c)$ in ultrahyperbolic space-time $\mathbbm{R}^{2,2}$ with coordinates $x^\mu=(w, z, \tilde{w}, \tilde{z})$ and metric $ds^2=2(dz d\tilde{z}-dw d\tilde{w})$, see App. \ref{app:notations} for details of our notation and conventions. We make this choice of space-time signature as it allows for the construction of real solutions to the ASDYM equations and for a straightforward reduction to the 2+1-dimensional integrable chiral model which motivates many of our considerations. The covariant derivative is defined in terms of the anti-Hermitian, Lie algebra valued gauge field $A_\mu\in \mathfrak{su}(N_c)$ as $D_\mu =\partial_\mu +A_\mu$ and the 
field strength is given by
\begin{align}
	F_{\mu \nu}\equiv [D_\mu, D_\nu]=\partial_\mu A_\nu-\partial_\nu A_\mu+[A_\mu, A_\nu]~.
\end{align}
 In order to define the anti-self-dual theory we introduce the Hodge star operator
\begin{align}
\ast F_{\mu \nu}=\frac{1}{2} \sqrt{|\det g|} \epsilon_{\mu \nu \rho \delta} F^{\rho \delta}
\end{align}
where $\epsilon_{\mu \nu \rho \delta}$ is the totally anti-symmetric Levi-Civita tensor with $\epsilon_{w\tilde{w}z\tilde{z}}=+1$ and $\det g$ is the determinant of the space-time metric. 
We define self-dual two-forms as those which satisfy $F=\ast F$ and anti-self-dual two-forms as those which satisfy $F=-\ast F$. Imposing anti-self-duality on the field strength implies 
\begin{align}
F_{wz}=F_{\tilde{w}\tilde{z}}=0~,~~~
F_{w\tilde{w}}-F_{z \tilde z}=0~.
\end{align}
In Minkowski signature these relations can only be satisfied by complex field strengths but there are real solutions in Euclidean and ultrahyperbolic spaces.
It is often useful to make use of spinor notation where every two-form  has a spinor equivalent which can be decomposed into two-index symmetric spinors,
$
F_{ABA'B'}=\phi_{AB}\epsilon_{A'B'}+\tilde{\phi}_{A'B'}\epsilon_{AB}
$,
where $\phi_{AB}$ describes the ASD part and $\tilde{\phi}_{A'B'}$ the SD part of the two-form. It is worthwhile to note that any solution to the SD or ASD conditions will automatically satisfy the Yang-Mills equation of motion as 
\begin{align}
D_\mu F^{\mu \nu}=\pm D_\mu (\ast F^{\mu \nu})=0
\end{align}
where the last equality is simply the Bianchi identity and so all solutions we consider will also be solutions of Yang-Mills in $\mathbbm{R}^{2,2}$.

\subsection{Lax representation}
The ASDYM equations, which we can write as
\begin{align}
[D_z, D_w]=0~, ~~~[D_{\tilde z}, D_{\tilde w}]=0~, ~~~[D_z, D_{\tilde z}]-[D_{w}, D_{\tilde w}]=0
\end{align} 
are equivalent to the commutativity of the Lax operators $L$ and $M$, where
\begin{align}
L=D_w-\zeta D_{\tilde z}~,~~~M=D_z-\zeta D_{\tilde w}~,
\end{align}
for every value of the complex spectral parameter $\zeta$. Thus the ASDYM equations can be viewed as the consistency conditions for the system of Lax equations
\begin{align}
	\label{eq:Lax1}
	L\psi(\zeta)=0~,~~~M\psi(\zeta)=0
\end{align}
where $\psi(\zeta)$ is a matrix-valued function of the space-time coordinates and the spectral parameter. Under gauge transformations, $A_\mu\to g A_\mu g^{-1}- (\partial_\mu g ) g^{-1}$, the Lax operators transform as 
$L\to gLg^{-1}$, $M\to g M g^{-1}$
and so solutions to the Lax equations transform as $\psi\to g \psi$. Introducing the two-spinor $\tilde{\lambda}^{A'}=(-\zeta, 1)$ and defining the operators $L_A=\tilde{\lambda}^{A'}D_{AA'}=(L,M)$ the consistency conditions 
\begin{align}
	[L_A,L_B]=\tilde{\lambda}^{A'}\tilde{\lambda}^{B'} F_{ABA'B'}=0
\end{align}
can be seen to immediately follow from the vanishing of the self-dual field strength component $\tilde{\phi}_{A'B'}$.

\subsection{Yang's formulation}
There is a convenient formulation of (anti-)self-dual Yang-Mills due to Yang \cite{Yang:1977zf}. The ASDYM equations guarantee the existence of the matrix-valued functions $h=\psi(\zeta=0)$ and $\tilde{h}=\psi(\zeta=\infty)$ satisfying the equations
\begin{align}
\partial_w h+A_w h =0~,& ~~\partial_z h+A_z h=0\nn\\
\partial_{\tilde w} {\tilde h}+A_{\tilde w}\tilde{h}=0~, & ~~\partial_{\tilde z} {\tilde h}+A_{\tilde z}\tilde{h}=0~
\end{align}
where under a gauge transformation $g$, we have that ${h}\to g h$ and $\tilde{h}\to g \tilde{h}$. If we assume that $h$ and $\tilde{h}$ are invertible, then we can define Yang's matrix 
$J={\tilde  h}^{-1} {h}$ which is invariant under gauge transformations and determines the gauge potentials via
\begin{align}
(\partial_{ w} J) J^{-1}&=
-{\tilde h}^{-1} A_{w} {\tilde h}-{\tilde h}^{-1}\partial_{ w} \tilde{h}	
\nn\\
J^{-1} (\partial_{\tilde{w}}J)&=h^{-1} A_{\tilde w} h+h^{-1}\partial_{\tilde w} h
\end{align}
and similarly for $z$ and $\tilde z$. The ASD equations then imply that $J$ satisfies Yang's equation
\begin{align}
\partial_{\tilde{w}}((\partial_{{w}} J) J^{-1} )-\partial_{\tilde z}(( \partial_{{z}} J )J^{-1})=0~.
\end{align}
A particularly simple gauge is the S-gauge 
$h=J$, $\tilde{h}=\mathbbm{1}$
such that 
\begin{align}
A_{\tilde{w}}=A_{\tilde z}=0~,~~~A_{w}=-(\partial_w J)J^{-1}~,~~~A_{z}=-(\partial_z J) J^{-1}~,
\end{align}
where the Lax operators become
\begin{align}
L=(\partial_w -(\partial_w J) J^{-1})- \zeta \partial_{\tilde z}  ~,~~~M=(\partial_z-(\partial_z J) J^{-1}) -\zeta \partial_{\tilde w} ~.
\end{align} 
This form of the ASDYM equations is a particularly suitable starting point for the Darboux solution generating method which naturally produces $J$. For our purposes, where we use the Zakharov approach, it is more direct to consider the so-called $K$-matrix formulation. In the S-gauge we have from the ASDYM equations that
\begin{align}
	\partial_{\tilde{z}}A_z-\partial_{\tilde w}A_w=0
\end{align}
which implies that there exists a $K$ such that $A_z=\partial_{\tilde w}K$ and $A_w=\partial_{\tilde z}K$. The remaining equation of motion is then 
\begin{align}
	\label{eq:Keq}
	\Box K-[\partial_{\tilde z}K, \partial_{\tilde{w}}K]=0~,
\end{align}
where we define 
\begin{align}
	\Box=\frac{1}{2} g^{\mu \nu} \partial_\mu \partial_\nu=\partial_{z}\partial_{\tilde{z}}-\partial_{w}\partial_{\tilde{w}}~.
\end{align}

\subsection{Scalar densities}
In order to characterise the classical solutions it is useful to consider gauge-invariant scalar quantities. One natural choice is the Yang-Mills Lagrangian density 
\begin{align}
	\mathcal{L}&=-\frac{1}{4} \mathrm{Tr}(F^{\mu\nu}F_{\mu\nu})
\end{align}
For the case of anti-self-dual gauge fields this is equivalent to the usual $\theta$-term proportional to Tr$(\ast F_{\mu \nu}F^{\mu \nu})$ and so is a total derivative. In our double-null coordinates we have that the Lagrangian density can be written as
\begin{align}
	\label{eq:YML_comp}
	\mathcal{L}=\mathrm{Tr}(F^2_{w\tilde{w}}-F_{w\tilde{z}}F_{z\tilde{w}})~.
\end{align}
Using the $K$-matrix form for the gauge potentials, the non-vanishing field strengths are 
\begin{align}
	F_{w \tilde{z}}=- \partial^2_{\tilde z} K~, ~~~F_{z \tilde{w}}=-\partial^2_{\tilde w} K~,~~~	F_{w \tilde{w}}=F_{z \tilde{z}}=- \partial_{\tilde w}\partial_{\tilde z} K~,
\end{align}
and so the Lagrangian density evaluated on solutions is
\begin{align}
	\mathcal{L}=  \text{Tr}\Big[(\partial_{\tilde w}\partial_{\tilde z} K)^2-(\partial^2_{\tilde z} K)(\partial^2_{\tilde w} K)\Big]~
\end{align}
which can be written as a divergence
\begin{align}
	\label{eq:YML_td}
	\mathcal{L}=  \partial_{\tilde{w} } \mathcal{J}_{\tilde{z}}-\partial_{\tilde z}\mathcal{J}_{\tilde{w}}
\end{align}
where 
\begin{align}
\mathcal{J}_{\tilde{a}}&=\frac{1}{2}\text{Tr}\big[\partial_{\tilde z}K \partial_{\tilde{a}}\partial_{\tilde w} K -  \partial_{\tilde w}K \partial_{\tilde{a}}\partial_{\tilde z} K\big]=\frac{1}{2}\text{Tr}[A_w \partial_{\tilde{a}} A_z-A_z \partial_{\tilde{a}} A_w]
\end{align}
are Chern-Simons currents. This total derivative form ensures that the integrated action is just a boundary term and vanishes for appropriate fall-off conditions. 
 As we will see, in general our solutions will have non-trivial densities but vanishing total action and, insofar as it is meaningful in split-signature, they are in the instanton-number zero sector. However, solutions can have isolated singular points corresponding to defects which give non-vanishing action. It is possible to write an action for which \eqref{eq:Keq} is the Euler-Lagrange equation \cite{Leznov:1986up, Leznov:1986mx, Parkes:1992rz}. 
 The equation of motion for $K$ also follows from the Chalmers-Siegel action \cite{Siegel:1992xp, Chalmers:1996rq}
  \begin{align}
  	\label{eq:CSact}
 		S_{ChS}=\mathrm{Tr} \int d^4x\, \Psi \left(\Box  K  -[\partial_{\tilde{z}} K, \partial_{\tilde{w}} K]\right)
 \end{align}
where $\Psi$ describes the opposite helicity gauge field. This action
   has in turn played an important role in the computation of perturbative Yang-Mills scattering amplitudes. 

 While the Yang-Mills Lagrangian density is a natural choice to characterise solutions it does not uniquely distinguish them, and can be somewhat involved to compute, so additional densities are desirable. The symmetric Belinfante energy-momentum tensor vanishes trivially at all space-time points for anti-self-dual solutions and so is not useful. We will instead introduce a $\tau$-function and the corresponding $\sigma$-density, $\sigma=\Box \log \tau$ which can be used to determine $\mathcal{L}$. 

\section{Dressing solutions}
\label{sec:ZS}
It is possible to construct non-trivial solitons for ASDYM by means of the dressing, or `proliferation of solutions',  method, due to Zakharov and collaborators e.g. \cite{zakharov1978relativistically, belinskii1978integration, zakharov1979integration, Forgacs:1980su}. Here we follow the treatment by Ward \cite{Hitchin:1999at} and for convenience we restrict to the case of $N_c=2$. Using the Lax equations \eqref{eq:Lax1},
we can construct an invertible $2 \times 2$ matrix $\psi$ from which we can extract the gauge potentials
\begin{align}
	\label{eq:IS_pot}
	A_w-\zeta A_{\tilde{z}}&=-(\big[\partial_w-\zeta \partial_{\tilde{z}}\big]\psi)\psi^{-1}\nn\\
	A_z-\zeta A_{\tilde{w}}&=-(\big[\partial_z-\zeta \partial_{\tilde{w}}\big]\psi)\psi^{-1}~.
\end{align}
To find an appropriate $\psi$ we assume that it is a rational function of $\zeta$, with simple poles at constant complex values, $\mu_1, \mu_2, \dots, \mu_N$, $\mu_i\neq \bar{\mu}_j$, with rank-one residues
\footnote{This assumption is in fact not needed for the SU$(2)$ case. From expanding $\psi(\zeta)\psi^{-1}(\zeta)=\mathbbm{1}$ near $\mu_k$ we have $M^k \psi^{-1}(\mu_k)=0$ so that when $\psi^{-1}(\mu_k)\neq 0$, $M^k$ has rank one. When  $\psi^{-1}(\mu_k)$ vanishes one then has $\psi(\bar{\mu}_k)=0$ and the pole can be removed by a space-time independent rescaling of $\psi$ so that
	\begin{align}
\hat{\psi}(\zeta)=\left(\frac{\zeta-\mu_k}{\zeta-\bar{\mu}_k}\right){\psi}(\zeta)
	\end{align} 
is regular at $\mu_k$. Alternatively, we can assume that  the determinant has only simple zeros and poles
	\begin{align}
		\text{det}\, \psi(\zeta)=\prod_k \frac{\zeta-\bar{\mu}_k}{\zeta-\mu_k}~.
	\end{align}
} 
and goes to a constant as $\zeta\to \infty$.  
We further impose that $\psi$ satisfies a generalised unitarity condition
\begin{align}
	\psi(x, \zeta)\psi^\dagger (x, \bar{\zeta})=\mathbbm{1}~
\end{align}
which will guarantee an anti-Hermitian gauge field and real solutions in $\mathbbm{R}^{2,2}$. 
Choosing an appropriate normalisation, $\psi$ has the form
\begin{align}
	\label{eq:psi_int}
	\psi=\mathbbm{1}+\sum_{k=1}^N \frac{M^k(x)}{\zeta-\mu_k}
\end{align}
where we can write $M^k=v^k\otimes u^{kT}$. The unitarity condition further implies that 
\begin{align}
v_a^{k'}=-(\Gamma^{-1})^{k'k}\bar{u}_a^k~~~\text{with}~~~
\Gamma^{kk'}=\frac{\lpangle {u}^{k}, u^{k'}\rpangle }{\bar{\mu}_k-\mu_{k'}}
\end{align}
where $\lpangle u^k, u^{k'}\rpangle =(u^k)^\dagger u^{k'}$.
$\Gamma$ is the $N\times N$ soliton matrix which, with our choice of reality conditions, is anti-Hermitian, $\Gamma^\dagger=-\Gamma$, and we require that it is invertible for all space-time points such that det$\,\Gamma\neq 0$. 

We can in fact provide the stronger result that det$(-i \Gamma)>0$ by noting that $-i \Gamma\succ 0$ is a positive matrix if we take $\mu_i$ distinct and Im$\,\mu_i>0$ for all $i=1, \dots, N$. This follows from $\int_0^\infty dt e^{i(\mu_{k'}-\bar{\mu}_{k})t}=\tfrac{i}{\mu_{k'}-\bar{\mu}_{k}}$ as we can write $-i \Gamma$ as a Gram matrix
\begin{align}
-	i\Gamma^{kk'}=\int_0^\infty dt  \left(e^{ i{\mu}_k t}{u}_b^k\right)^\dagger \left(e^{i{\mu}_{k'} t}{u}_b^{k'}\right)\equiv\langle \langle \tilde{u}^k, \tilde{u}^{k'}\rangle\rangle
\end{align}
with $\tilde{u}^k=e^{i\mu_k t}u^k$ so that $\vec{w}^\dagger (-i\Gamma)\vec{w}=\|\tilde{u}\cdot \vec{w}\|^2> 0$ for $\vec{w}\neq 0$, where the inequality is strict  for $u^k\neq 0$ due to linear independence of the $\tilde{u}^k$ which is guaranteed by assuming the $\mu_k$ distinct. As a result, if we define the quantity 
\begin{align}
	\label{eq:tau}
	\tau=\prod_{k=1}^N (\bar{\mu}_k -\mu_k )\det (\Gamma)
\end{align}
it is positive and we may take the logarithm.  We will use this to characterise our soliton solutions below and show that it can be used to compute invariant quantities such as the Yang-Mills Lagrangian density. For combinations of poles both above and below the real axis, the soliton matrix $\Gamma$ remains non-singular and the product of Im$\, \mu_k$ factors ensures positivity.\footnote{Ordering the entries by Im$\,\mu_k$ we can write \begin{align} -i \Gamma=\begin{pmatrix} H_{++} & H_{+-} \\
H^\dagger_{+-} & H_{--} \end{pmatrix}
\end{align}
where $H_{++}\succ 0$ and $H_{--}\prec 0$ by the argument above. Using $\det (-i \Gamma)=\det H_{++} \det S$, where the Schur complement is $S=H_{--}-H_{+-}^\dagger H_{++}^{-1} H_{+-}$, and the fact that $S\prec 0$ as
\begin{align}
	w^\dagger S w=w^\dagger H_{--} w - \| H_{++}^{-1/2} H_{+-}w\|^2< 0
\end{align}	
we have that 
\begin{align}
	\prod_{k=1}^N (\bar{\mu}_k -\mu_k )\det (\Gamma)=2^N \prod_{k=1}^N  |\text{Im}\mu_k| \det H_{++}\det (-S)>0~.
\end{align}}

In order that $\psi$ defines gauge potentials independent of $\zeta$, we must additionally impose the vanishing of poles on the RHS of \eqref{eq:IS_pot}. It is useful to introduce coordinates associated to each pole
\begin{align}
\ell_k=\mu_k w+ \tilde{z}~, ~~~m_k=\mu_k z+ \tilde{w}~,~~~\bar{\ell}_k=\bar{\mu}_k w+ \tilde{z}~, ~~~\bar{m}_k=\bar{\mu}_k z+ \tilde{w}~
\end{align} 
or using spinor notation
\begin{align}
	(\ell_k, m_k)^{A}=\tilde{\lambda}_{A',k}x^{AA'}
\end{align}
with $\tilde{\lambda}_k=(1,\mu_k)$. 
Hence, defining the anti-holomorphic derivatives
\begin{align}
\partial_{\bar{\ell}_k}=-\frac{1}{\kappa_k}(\partial_w-\mu_k \partial_{\tilde{z}})~,~~~ \partial_{\bar{m}_k}=-\frac{1}{\kappa_k}(\partial_z-\mu_k \partial_{\tilde{w}})~~~\text{with}~~~\kappa_k=\mu_k-\bar{\mu}_k
\end{align}
we have
$\partial_{\bar{\ell}_k} u^k=\alpha u^k$, $\partial_{\bar{m}_k} u^k=\beta u^k$
and, given that $\psi$ depends on $u^k$ projectively, we can choose $\alpha=\beta=0$ such that
 \begin{align}
\partial_{\bar{\ell}_k} u^k=\partial_{\bar{m}_k}u^k=0~.
\end{align}
  Thus $u^k$ is only a function of the combinations $\ell_k$ and $m_k$ and similarly, $\bar{u}^k$ is only a function of $\bar{\ell}_k$ and $\bar{m}_k$. 

The basic data to specify a solution are the set of complex poles $\{\mu_k\}$ and the polarisation vectors $u^k$ which are defined up to a global U$(2)$ transformation. Furthermore, we may separately rescale each $u^k\to \lambda_k u^k$ by any holomorphic function $\lambda_k(\ell_k, m_k)$ such that each $u^k$ should be thought of as a homogeneous coordinate of $\mathbbm{C}P^1$. The $\tau$-function is consequently arbitrary up to rescalings
\begin{align}
	\label{eq:tau_rescale}
	\log \tau\to \log \tau+\sum_{k=1}^N\log |\lambda_k(\ell_k, m_k)|^2
\end{align}
but is invariant under global SU$(2)$ gauge transformations. Lorentz transformations in split signature, SO$_0(2,2)\simeq($SL$(2, \mathbbm{R})\times$SL$(2, \mathbbm{R}))/\mathbbm{Z}_2$, act as $x\to g \cdot x\cdot \tilde{g}^T$. Equivalently, the primed transformation acts on the spinor $\tilde{\lambda}_{A'} \to \tilde{\lambda}_{B'} \tilde{g}^{B'}_{A'}$
\begin{align}
	 (1, \mu_k)\to (a+\mu_k c)(1,\mu_k^g )~,~~~\text{with}~~~\mu_k^g=\frac{b+\mu_k d}{a+\mu_k c}, ~~~ ad-bc=1~
\end{align}
such that the spectral parameter transforms by a M\"obius transformation and there is a corresponding rescaling and rotation of the doublet $(\ell_k, m_k)\to (a+\mu_k c) g\cdot (\ell_k, m_k)$. 

To determine the gauge fields from $\psi$, we use the fact that in \eqref{eq:IS_pot} the RHS has no poles and is bounded as $\zeta\to \infty$. Thus the terms linear in $\zeta$ on the LHS vanish and so we immediately have $A_{\tilde z}=A_{\tilde w}=0$. From the constant terms we further have that  
\begin{align}
	A_w=\partial_{\tilde z} K~,~~~A_z= \partial_{\tilde w} K~
\end{align}
where $K=\sum_{k=1}^N M^k$ which is essentially the $K$-matrix formulation of ASDYM. 

The solutions constructed in this way can also be related to Yang's formulation by setting the spectral parameter to zero and making the identification
\begin{align}
	J_{ab}=\psi_{ab}(\zeta=0)=\mathbbm{1}_{ab}+ {\bar{u}}_a   (\Gamma^{T})^{-1}D^{-1}u_b  ~, ~~~\text{det}\, J=\prod_k \frac{\bar{\mu}_k}{\mu_k}=\frac{\text{det}\, \bar{D}}{\text{det}\, D}
\end{align}
where   we have introduced the diagonal $N\times N$ matrix $D=\text{diag}(\mu_1,\dots,\mu_N)$, and $\bar{u}_a$ and $u_b$ are to be regarded as row and column vectors in the $N$-dimensional soliton space. This expression naturally leads to the appearance of quasideterminants \cite{gel1991determinants, gel1992theory, gelfand2005quasideterminants} which have previously been utilised in the construction of soliton solutions in ASDYM \cite{Nimmo:2000aj, Huang:2021ddb, Hamanaka:2021qsk, Huang:2022mqh}. In particular, introducing the matrix $U=u^k_a\in \mathbbm{C}^{2\times N}$, $J$ can be written as a quasideterminant,
\begin{align}
	J=\begin{vmatrix} D\Gamma^T & {U}^T
		\\ -\bar{U} & \fbox{$\mathbbm{1}$}
	\end{vmatrix}~.
\end{align}
This has the form of the Gramian solutions discussed in \cite{Li:2025dlc, Huang:2026vvw} which in turn are related to the Cauchy matrix solutions of \cite{Li:2021kxj, Li:2022fum}. This is, in essence, a consequence of the fact that the soliton matrix $\Gamma$ is a Cauchy-Gram type matrix satisfying the Sylvester equation 
\begin{align}
	\bar{D} \Gamma- \Gamma D = \bar{u}_b \otimes u^T_b~.
\end{align}

\section{One-pole solutions}
Solution sectors are labelled by the number of poles of the matrix function $\psi$ though it is worth noting that this need not equal the number of localised lumps, walls or other structures in space-time.  Solutions with a single pole, which we refer to as one-soliton solutions, are given by
\begin{align}
	\psi=\mathbbm{1}+\frac{\kappa}{\zeta-\mu} P~,~~~ P_{ab}=\frac{\bar{u}_a u_b}{\|u\|^2}~
\end{align}
where $P$ is a $2\times 2$ projection matrix, $P^2=P$, and $\kappa=\mu-\bar{\mu}$. 
If we additionally use the freedom to rescale $u$ we can choose
\begin{align}
	u=\begin{pmatrix} f(\ell, m)\\ 1\end{pmatrix}
\end{align}
where $f$ is a holomorphic function of $\ell=\mu w+ \tilde{z}$ and $m=\mu z+ \tilde{w}$ which results in
\begin{align}
	P=\frac{1}{1+|f|^2}\begin{pmatrix} |f|^2 & \bar{f}\\ f & 1 \end{pmatrix}~.
\end{align}
The Lagrangian density in the one-soliton sector can be written in a particularly simple form
\begin{align}
	\mathcal{L}=-\Box \sigma
\end{align}
where $\Box=\partial_z \partial_{\tilde{z}}-\partial_w\partial_{\tilde{w}}=\kappa(\partial_m\partial_{\bar{\ell}}-\partial_\ell \partial_{\bar{m}})$ and
\begin{align}
	\sigma=\Box\log\tau~,~~~\tau= -\kappa \det \Gamma~.
\end{align}
This is the simplest case of the general result so we have included the determinant which, though trivial for a $1\times 1$ matrix, will be relevant for $N>1$. 
As the one-soliton solution is given by a map $f:\mathbbm{R}^{2,2}\to \mathbbm{C}P^1$ it is naturally given in terms of the Fubini-Study structure. In particular, the logarithm of $\tau$ is, up to constants, the K\"ahler potential
\begin{align}
 \mathcal{K}=	 \log (1+|f|^2)~,
\end{align}
and $\sigma$ is related to the pull-back of the Fubini-Study two-form 
\begin{align}
	\label{eq:sigma_def}
	\sigma=	\kappa \frac{f_m\bar{f}_{\bar{\ell}}-f_{\ell}\bar{f}_{\bar{m}}}{(1+|f|^2)^2}~.
\end{align}
Finally, we can give an explicit expression for the Lagrangian density in this sector
\begin{align}
	\mathcal{L}=\kappa^2  (2 g_{m \ell} \bar{g}_{\bar{\ell}\bar{m}}-g_{\ell \ell} \bar{g}_{\bar{m}\bar{m}}-g_{mm} \bar{g}_{\bar{\ell}\bar{\ell}})+2 \sigma^2
\end{align}
where $g_{\alpha \beta}$, $\alpha=\ell, m$, is related to the covariant Hessian,
\begin{align}
		g_{\alpha \beta}=\frac{1}{\|u\|^2}\left(f_{\alpha\beta}-2\frac{f_\alpha f_\beta}{\|u\|^2} \bar{f}\right)~, ~~~	\bar{g}_{\bar{\alpha}\bar{\beta}}=\frac{1}{\|u\|^2}\left(\bar{f}_{\bar{\alpha}\bar{\beta}}
		-2\frac{\bar{f}_{\bar{\alpha}} \bar{f}_{\bar \beta}}{\|u\|^2} {f}\right)~.
	\end{align}

\subsection{Lump-solitons}
\label{sec:lumpsol}
If we choose $f$ to be a polynomial function of holomorphic coordinates $\ell$ and $m$ we find lump solitons. The degree of the polynomial acts as a discrete label of the solution. For the simplest case 
\begin{align}
	f(\ell, m)=a_\ell \ell + a_m m~, ~~~a_\ell, a_m \in \mathbbm{C}
\end{align}
we find the Lagrangian density simplifies to 
\begin{align}
	\mathcal{L}=2(p\cdot \bar{p})^2 \frac{1-2 |f|^2}{(1+|f|^2)^4}
\end{align}
where we have introduced a null complex soliton momentum $p_z=\mu a_m$, $p_w=\mu a_\ell$, $p_{\tilde{z}}=a_\ell$, $p_{\tilde{w}}=a_m$ which satisfies $p^2=0$ with the split signature metric. 
\begin{figure}
	\centering
	\includegraphics[scale=0.5]{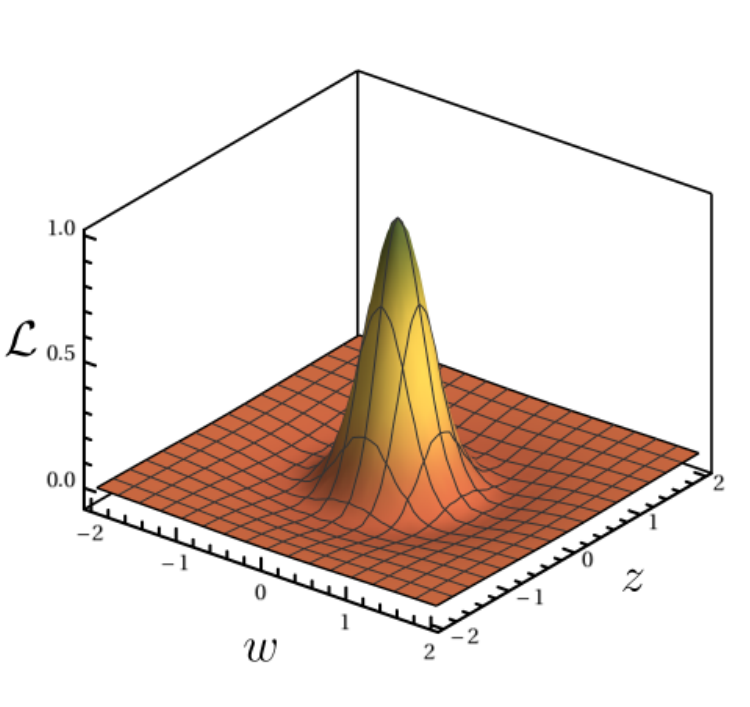}~~~	\includegraphics[scale=0.5]{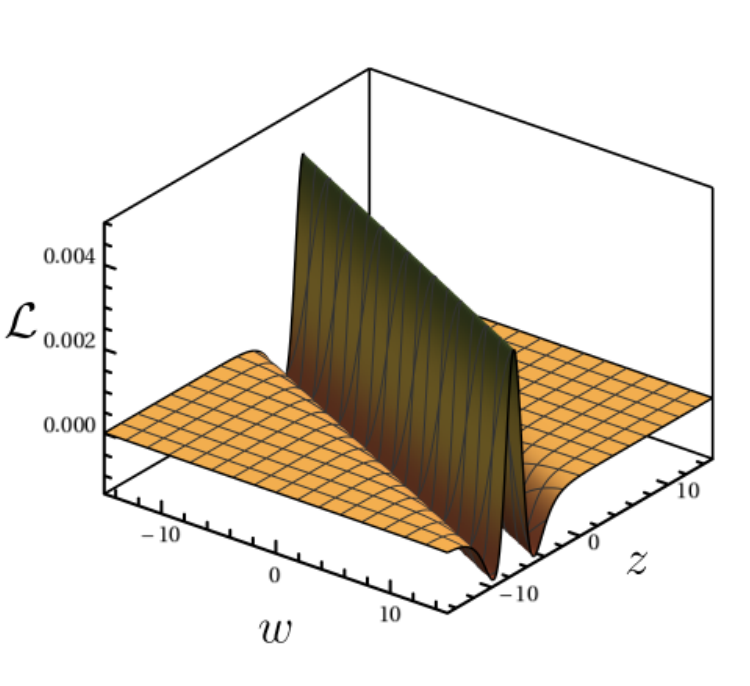}
	\caption{Left: Lagrangian density for lump soliton with $\mu=i$, $\tilde{z}=\tilde{w}=0$, $a_\ell=i, a_m=1$. Right: Lagrangian density for extended wave soliton with $\mu=i/2$, $c=1$, $\tilde{z}=\tilde{w}=0$, $a_\ell=i, a_m=(1+i)/2$.}
	\label{fig:1lumpLag}
\end{figure}
We can see that $\mathcal{L}$ has a maximum for $|f|=0$ and goes to zero for large $|f|^2$, see Fig. \ref{fig:1lumpLag}. It is not positive definite and is negative for $|f|^2>1/2$. Indeed, computing the action by integrating over all space-time gives zero. As the solutions are ASD this is consistent with being in the zero-instanton sector. 
Alternatively, the solution is characterised by the $\tau$-function proportional to $1+|f|^2$ which gives rise to the scalar 
\begin{align}
	\sigma=\frac{p\cdot \bar{p}}{(1+|f|^2)^2}
\end{align}
whose absolute value also has a maximum at $|f|=0$ and then monotonically asymptotes to zero as $|f|\to \infty$. To be clear, we refer to these solutions as lumps as they are analogous to the localised, rational solutions found in the KP equation and by Ward \cite{Ward:1988ie, Leese:1989hj} for the 2+1 chiral model. They are however localised on two-dimensional surfaces defined by the complex equation $f=0$. Higher-degree polynomials will have more complicated profiles, typically with several such localised lumps. 

We can generalise such solutions to rational functions
\begin{align}
	f(\ell, m)=\frac{a_\ell \ell +a_m m}{b_\ell \ell + b_m m}
\end{align}
which similarly describe lump-solitons. However, in this case there are space-time points where $u=0$, and where the zero cannot be removed by a holomorphic rescaling, which give rise to singular gauge fields. For example, taking $u=(\ell, m)$ and so $f=\ell/m$ the polarisation vanishes at the point $\ell=0$, $m=0$. In this simple case we have that 
\begin{align}
\sigma=\kappa \frac{\bar{\ell} m- \ell \bar{m}}{(|\ell|^2+|m|^2)^2}~,~~~\mathcal{L}=2\kappa^2 \frac{(|\ell|^2+|m|^2)^2+ 3 (\ell \bar{m}-\bar{\ell} m)^2}{(|\ell|^2+|m|^2)^4} ~.
\end{align}

This Lagrangian density is singular at the origin and, being homogeneous of degree $-4$, its integral over space-time is naively divergent at both the origin and at infinity, such that computing the action requires a prescription. A specific choice starts by converting to complex coordinates $(\ell,\bar{\ell}, m,\bar{m})$ and then integrating over the level sets of $\tau$ i.e. the spheres $|\ell|^2+|m|^2=r^2$. The integration of $\mathcal{L}$ over each sphere vanishes. However, writing $\mathcal{L}=-\Box\sigma=\partial_\mu V^\mu$ with $V^\mu=-\frac{1}{2}g^{\mu\nu}\partial_\nu\sigma$, and noting that $\sigma$ is homogeneous of degree $-2$, the flux
\begin{align}
	\oint_{\tau=R^2}V^\mu\, d\Sigma_\mu=\frac{4\pi^2}{3}
\end{align}
is independent of $R$.  This can be given an alternative interpretation by noting that $\tau_\epsilon=|\ell|^2+|m|^2+\epsilon^2$ is the $\tau$-function of the smooth one-pole SU$(3)$ solution with $u=(\ell,m,\epsilon)$, which reduces to the SU$(2)$ lump as $\epsilon\to0$. Computing the angular integral of the corresponding Lagrangian density over the sphere of radius $r$  and then integrating over $r$ gives
\begin{align}
	S=\int d^4x\,\mathcal{L}_\epsilon=\frac{4\pi^2}{3}
\end{align}
for every $\epsilon$.  Thus we see that with an appropriate choice of regularisation we can have rational solutions decorated by finite-action defects. We will see that these defects can also occur when we consider wall-type solitons.

\subsection{Extended wave solitons}
The ASDYM analogue of the extended wave solitons constructed in \cite{Leese:1989hj}  can be found by taking $f(\ell, m)=c~ \text{exp}(a_\ell \ell + a_m m )$. This is also the soliton wall solution found in \cite{Hamanaka:2020pyc} based on the iterative Darboux construction \cite{Nimmo:2000aj}, see also \cite{deVega:1987dj} for earlier work. The Lagrangian density has the form
\begin{align}
\mathcal{L}=-(p\cdot \bar{p})^2\frac{|f|^2(1-4 |f|^2+|f|^4)}{(1+|f|^2)^4}
\end{align}
which is the same result as found in  \cite{Hamanaka:2020pyc}. This can be made slightly clearer by 
introducing the argument 
\begin{align}
	\label{eq:ext_wv_1}
\phi=k\cdot x+\log |c|=\text{Re}(a_\ell \ell +a_m m)+\log |c|
\end{align}
 where $k$ is a
 real, massive, wave-vector
\begin{align}
	k_w=\text{Re}( a_\ell \mu), ~~~k_z= \text{Re}(a_m \mu)~,~~~k_{\tilde w}= \text{Re}(a_m)~,~~~ k_{\tilde z}=\text{Re}(a_\ell)~,
	\end{align}
with $k^2=i\kappa \text{Im}(a_m\bar{a}_\ell)$. For this solution we have $\tau=(1+e^{2\phi})$, for the scalar $\sigma$
\begin{align}
	\sigma&=p\cdot \bar{p} \frac{|f|^2}{(1+|f|^2)^2}
	=\frac{k^2}{2} \text{sech}^2 \phi ~,
\end{align} 
and we can rewrite the Lagrangian density as
\begin{align}
	\mathcal{L}= \frac{1}{2}k^4 (3 \text{sech}^4 \phi-2 \text{sech}^2 \phi)~.
\end{align}
 This solution describes a codimension-one soliton whose profile localises on the three-dimensional hyperplane defined by $\phi=0$, see Fig. \ref{fig:1lumpLag}, right panel. 

\subsection{Two-phase wall solitons}
\label{sec:two_phase_wall}
\begin{figure}
	\centering
	\raisebox{-0.5\height}{\includegraphics[scale=0.5]{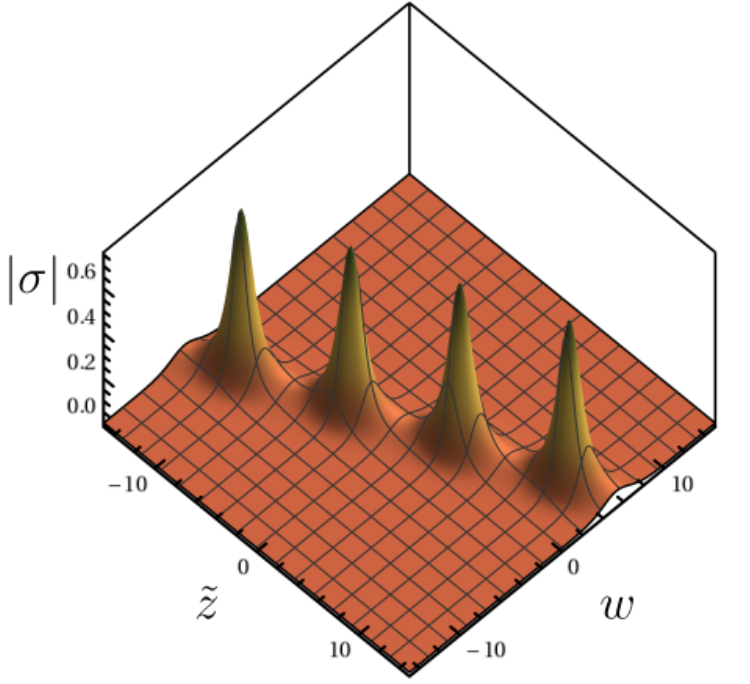}}\hspace{2cm}\raisebox{-0.5\height}{\includegraphics[scale=0.25]{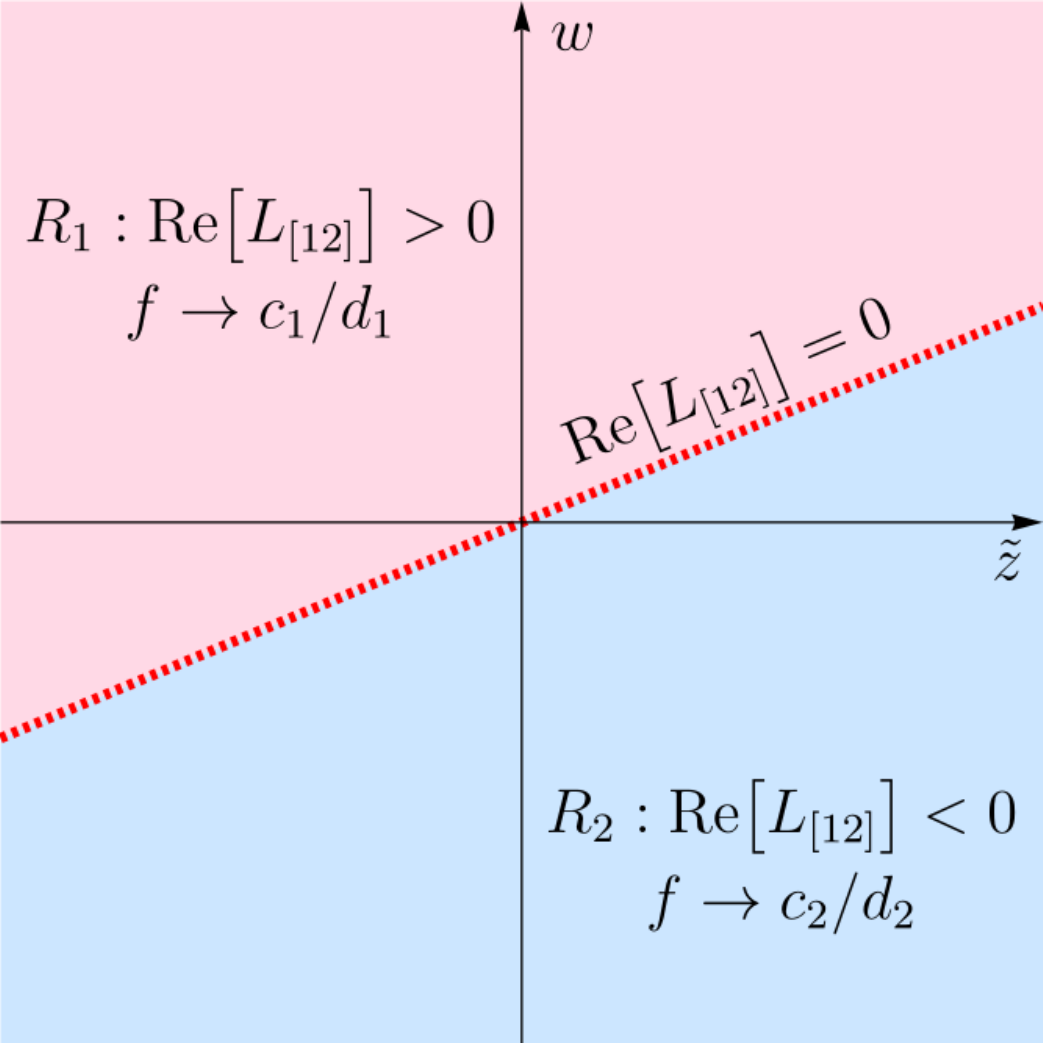}}
	\caption{Left: $|\sigma|$ for two-phase modulated extended  wave soliton with $\mu=i$, $m={z}=\tilde{w}=0$, $a_{\ell,1}=e^{-i\pi/4}$, $a_{\ell,2}=1$, $\lpangle q_1, q_2 \rpangle=\bar{c}_1 c_2 +\bar{d}_1 d_2=1/2$. Right: Regions of space-time defined by asymptotics of two-phase solution. }
	\label{fig:2phase1Ew}
\end{figure}

 We can slightly generalise this by considering the two-phase (but still one-pole or one-soliton) solution
\begin{align}
f(\ell,m)=\frac{c_1 +c_2 e^{\theta_2-\theta_1}}{d_1+d_2 e^{\theta_2-\theta_1}}
\end{align}
where $\theta_i=p_i\cdot x$ and
\begin{align}
	p_{w,i}=a_{\ell,i} \mu ~,~~~ p_{z,i}= a_{m,i} \mu~,~~~p_{\tilde w,i}= a_{m,i}~,~~~ p_{\tilde z,i}=a_{\ell,i}~,~~~i=1,2
\end{align}
are both complex null momenta with respect to the $2+2$ signature metric
$
	p_i^2=0$.
As $p$ is null we can write it as a product of spinors, $(p_i)_{AA'}=\lambda_{A,i} \tilde{\lambda}_{A'}$ with $\lambda_{A,i}=(a_{\ell,i}, a_{m,i})$, $\tilde{\lambda}_{A'}=(1, \mu)$. 
This solution can equivalently be written in terms of the polarisation vector
\begin{align}
	u=q_1 E_1 +q_2 E_2
\end{align}
where we have introduced the notation $E_i=e^{\theta_i}$ and the constant colour vectors
\begin{align}
	q_1=\begin{pmatrix} c_1 \\ d_1\end{pmatrix} ~,~~~q_2=\begin{pmatrix} c_2 \\ d_2\end{pmatrix}~
\end{align}
where we assume $\det( q_1 , q_2) \neq0$. The $\tau$-function is given by 
\begin{align}
	\label{eq:tau_two_phase}
\tau=\|u\|^2=2 \|q_1\|\|q_2\| |E_1 E_2| \left(\cosh \text{Re}(L_{[12]})+\cos \chi_{12} \cos \Phi_{12}\right)
\end{align}
where $\|q_i\|^2=q_i^\dagger q_i=\lpangle q_i, q_i\rpangle$, and it is convenient to introduce notation for the complex wall function
\begin{align}
	L_{[12]}=\theta_1-\theta_2+\log\frac{\|q_1\|}{\|q_2\|}
\end{align}
where the angle and phase are defined as
\begin{align}
\cos \chi_{12} e^{i\phi_0}=\frac{\lpangle q_1, q_2\rpangle }{\|q_1\|\|q_2\| }~,~~~\Phi_{12}=\phi_0-\text{Im}\,L_{[12]}~.
\end{align}
The solution splits space-time into two parts 
\begin{align}
R_i=\{x^\mu: \rho_i(x)>\rho_j(x),\, i\neq j\}
\end{align}
where $\rho_i=\text{Re}\,\theta_i+\log \|q_i\|$. 
In the asymptotic region of $R_1$, where $\rho_1\gg \rho_2$, we have that $f\to c_1/d_1$ while in $R_2$ we have $f\to c_2/d_2$. 
The soliton profile is non-trivial along the wall defined by the condition $\text{Re}\,L_{[12]}=0$ where the two exponents $\theta_1$, $\theta_2$, balance and so we denote this wall by $[12]$. We can compute the scalar $\sigma$ for this solution
\begin{align}
	\sigma=\Box\log \tau=\frac{p_{12}\cdot \bar{p}_{12}\sin^2 \chi_{12}}{4\left(\cosh \text{Re}(L_{[12]})+\cos \chi_{12} \cos \Phi_{12}\right)^2}
\end{align}
where $p_{12}=p_1-p_2$. We can see that $|\sigma|$ is maximised along Re$\,L_{[12]}=0$. When $\chi_{12}=\pi/2$, that is when $q_1 \perp q_2$, the wall is uniform but for other values the wall is modulated by $\cos \Phi_{12}$ and vanishes at $\chi_{12}=0$ when $q_1 \parallel q_2$, see Fig. \ref{fig:2phase1Ew} for an example. 

Thus we see that the data which encodes the profile of the wall is the real $\mathbbm{R}^{2,2}$ momentum vector, $k_{[12]}^\mu=\text{Re}(p^\mu_1-p_2^\mu)$, normal to the wall, the ratio $\|q_1\|/\|q_2\|$ which determines the position of the wall profile, the angle $\chi_{12}$ related to the colour Bloch sphere which determines the maximum wall amplitude, and  Im$\,(p^\mu_1-p_2^\mu)$ which together with the phase $\phi_0$ fixes the modulations along the wall.

\subsection{Y-solitons}
\begin{figure}
	\centering
		\raisebox{-0.5\height}{\includegraphics[scale=0.5]{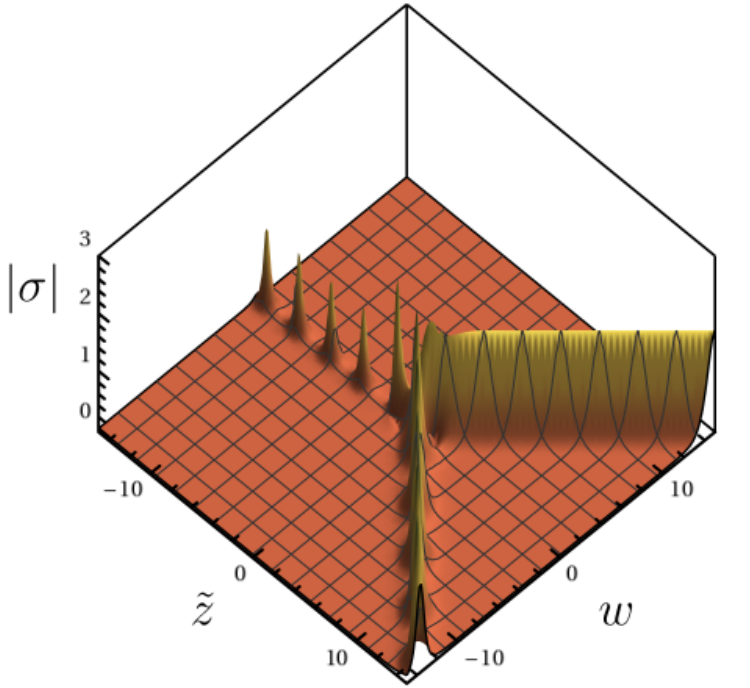}}\hspace{2cm}	\raisebox{-0.5\height}{\includegraphics[scale=0.25]{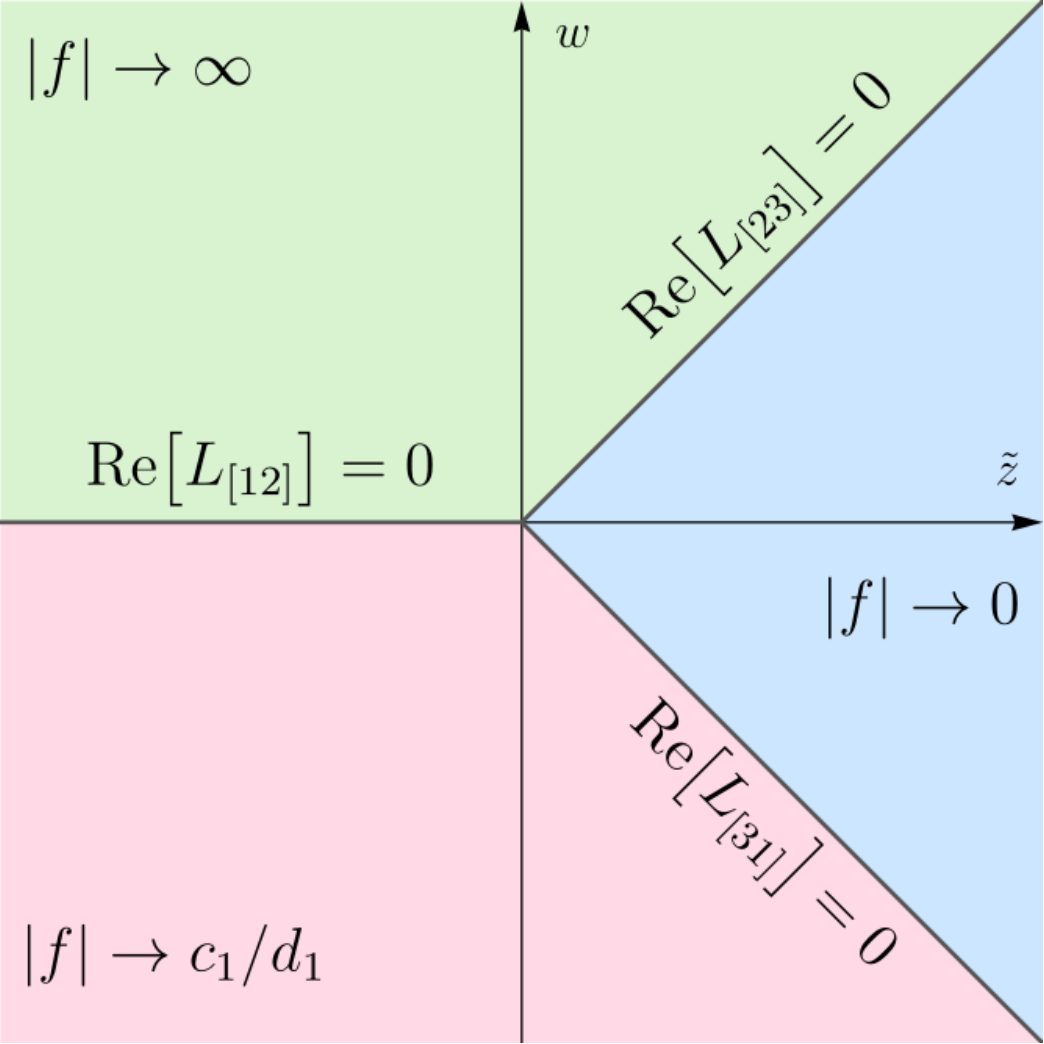}}
	\caption{Left: $|\sigma|$ for extended wave Y-soliton with $\mu=i$, $m={z}=\tilde{w}=0$, $a_{\ell,1}=e^{i\pi/2}$, $a_{\ell,2}=e^{-i\pi/2}$, $a_{\ell,3}=1$, $a_{m,1}=e^{-i\pi/3}$, $a_{m,2}=e^{i\pi/\sqrt{7}}$, $a_{m,3}=e^{-i\pi/\sqrt{5}}$, $c_1=d_1=2$, $c_2=d_2=1$. Right: Regions of the $w-\tilde{z}$ plane where each of the exponentials dominates and boundaries separating $|f|\to 0$ (blue), $|f|\to \infty$ (green), $|f|\to c_1/d_1$ (pink) regions corresponding to $[12]$, $[23]$, $[31]$ soliton arms.  }
	\label{fig:Ysol}
\end{figure}
A further generalisation can be found by considering rational functions of three exponentials 
\begin{align}
	f(\ell,m)=\frac{c_1 e^{\theta_1}+c_2 e^{\theta_2}}{d_1 e^{\theta_1}+d_2 e^{\theta_3}}~,
\end{align}
or equivalently by taking the polarisation vector to be a sum of three exponentials
\begin{align}
	\label{eq:uYex}
	u(\ell, m)=\sum_{i=1}^3 q_i E_i=\begin{pmatrix} c_1 \\ d_1 \end{pmatrix}e^{\theta_1} +
	 \begin{pmatrix} c_2 \\ 0 \end{pmatrix}e^{\theta_2}+
	 \begin{pmatrix} 0 \\ d_2 \end{pmatrix} e^{\theta_3}~.
\end{align}
 There are now three asymptotic regions: i) where Re$(\theta_1)$ dominates and $f\sim c_1/d_1$ ii) Re$(\theta_2)$ dominates and $f\sim \infty$ iii) Re$(\theta_3)$ dominates and $f\sim 0$. 
 The soliton profile is again non-vanishing along the boundaries between regions where different exponentials are dominant, see Fig. \ref{fig:Ysol}.
 For this solution we have that
\begin{align}
	\tau=\|u\|^2=\sum_{i,j=1}^{3}G_{ij} \bar{E}_i E_j
\end{align}
with
\begin{align}
	A=\begin{pmatrix} q_1 & q_2 & q_3 \end{pmatrix}~,~~~
	G=A^\dagger A=\begin{pmatrix} \bar{c}_1 c_1+\bar{d}_1 d_1 & \bar{c}_1 c_2 & \bar{d}_1 d_2 \\
		\bar{c}_2 c_1 & \bar{c}_2 c_2 & 0 \\
		\bar{d}_2 d_1 & 0 & \bar{d}_2 d_2 \end{pmatrix}~.
\end{align}
We use the notation $[ij]$ to denote the soliton wall separating the regions where the exponentials $E_i$ and $E_j$ are dominant. If we consider the $[12]$ wall, where $E_3$ is exponentially small, we can approximate $u\simeq q_1 E_1+ q_2 E_2$ and so 
\begin{align}
	\frac{\|u\|^2}{ \|q_1\|\|q_2\| |E_1| |E_2|}\simeq\left(2 \cosh \text{Re}[L_{[12]}]+2 \frac{\text{Re}(q_1^\dagger q_2 \bar{E}_1 E_2)}{\|q_1\|\|q_2\| |E_1| |E_2|} \right)~.
\end{align}
This gives rise to a non-trivial profile for $\sigma$ where the potentially exponentially large term  is minimised  along $\text{Re}[L_{[12]}]=0$. The codimension-one walls are more generally defined by the conditions
\begin{align}
	\text{Re}\big[L_{[12]} \big]=0~,~~~\text{Re}\big[L_{[23]} \big]=0~,~~~\text{Re}\big[L_{[31]} \big]=0
\end{align}
where again
\begin{align}
	L_{[ij]}=\theta_{i}-\theta_{j}+\log\frac{ \|q_i\|}{\|q_j\|}~.
\end{align}
 Writing $\theta_i=p_i\cdot x$, $i=1,2,3$, we can associate to each wall, $[ij]$,  wave-vectors $k^\mu_{[ij]}=$Re$(p^\mu_i-p^\mu_j)$ which then trivially satisfy the conservation, or resonance, condition,
\begin{align}
	k^\mu_{[12]}+k^\mu_{[23]}+k^\mu_{[31]}=0~.
\end{align} 
Moreover, we see that it is guaranteed that the three walls will intersect along a single two-surface as 
\begin{align}
	L_{[12]}+L_{[23]}+L_{[31]}=0~
\end{align}
so that at the intersection of two walls e.g. $\text{Re}[L_{[12]}]=\text{Re}[L_{[23]}]=0$ we automatically find $\text{Re}[L_{[31]}]=0$. 

For three phases, or indeed more, it is generically possible to find space-time points where $u=0$ and $\tau=0$. For the choice of parameters \eqref{eq:uYex}, this occurs at
\begin{align}
	\theta_1(\ell^\ast, m^\ast)-\theta_2(\ell^\ast, m^\ast)=\log\left(\frac{-c_2}{c_1}\right)+2\pi i n_{12}~,~~~\theta_1(\ell^\ast, m^\ast)-\theta_3(\ell^\ast, m^\ast)=\log\left(\frac{-d_2}{d_1}\right)+2\pi i n_{13}
\end{align}
which defines a lattice of points, $(n_{12}, n_{13})\in \mathbbm{Z}^2$. Near these points  we can approximate the solution by a rational function of the form
\begin{align}
	f\simeq \mathcal{N} \frac{a_{\ell,12} (\ell-\ell^\ast)+a_{m,12} (m-m^\ast)}{a_{\ell,13} (\ell-\ell^\ast)+a_{m,13} (m-m^\ast)}  
\end{align}
which is the finite-action version of the lump soliton, see Sec. \ref{sec:lumpsol}. These defects do not lie on the arms of the soliton but must lie close to the interaction region where the arms join. Thus each Y-soliton, and other higher-phase solutions, can be decorated by lump-like defects where $\tau=0$. For configurations that are three-dimensional, that is when $u$ is a function of just a single complex combination of $\ell$ and $m$, there will be no such zeros and correspondingly no such singular points are found in dimensionally reduced models such as the integrable chiral model. 

The form of the Y-soliton naturally suggests a connection with the interaction of three plane waves, that is the three-point scattering amplitude in ASDYM. We can make this connection slightly more concrete by  taking the momentum spinors, $\lambda_i=(a_{\ell, i}, a_{m, i})$, as points in $\mathbbm{Z}^2$. In this case the wall normals are real and null:  $k_{[ij]}=(\lambda_i-\lambda_j)\text{Re}(\tilde{\lambda})$ and $k_{[ij]}^2=0$. This implies that  the single-wall solitons have strictly vanishing profile,  however it leads to a simple geometric picture which is perhaps still worth highlighting. The polarisation vector
\begin{align}
	u=\sum_{i=1}^3 q_i e^{a_{\ell, i}\ell +a_{m, i}m}
\end{align}
can be viewed as a Laurent polynomial in variables $e^{\ell}$ and $e^{m}$ and the convex hull of the lattice points $\lambda_i$ in $\mathbbm{R}^2$ is the Newton polygon of $u$ \cite{sturmfels1998polynomial}
\begin{align}
	\text{New}(u)=\text{conv}\{\lambda_1, \lambda_2, \lambda_3\}~.
\end{align}
The oriented area of the Newton polygon, in this case a triangle, is 
\begin{align}
	A_{123}=\frac{1}{2}(\langle \lambda_1,\lambda_2\rangle +\langle \lambda_2,\lambda_3\rangle +\langle \lambda_3,\lambda_1\rangle)
\end{align}
where $\langle\lambda_{i},\lambda_j\rangle=a_{\ell,i}a_{m,j}-a_{\ell,j}a_{m,i}$ are the usual spinor brackets. If we define spinors for the soliton momenta $k_{[12]}=\pi_1 \text{Re}(\tilde{\lambda})$, $k_{[23]}=\pi_2\text{Re}(\tilde{\lambda})$, $k_{[31]}=\pi_3\text{Re}(\tilde{\lambda})$ then we have $\langle \pi_1,\pi_2\rangle =\langle \pi_2,\pi_3\rangle =\langle \pi_3,\pi_1\rangle =2A_{123}$. Consequently, the triangle's oriented area is given by
\begin{align}
A_{123}	=\frac{\langle \pi_1,\pi_2\rangle^3}{2\langle \pi_2,\pi_3\rangle \langle \pi_3,\pi_1\rangle }
\end{align}
which has the same form as the kinematic part, that is the colour-ordered and momentum $\delta$-function stripped, of the MHV amplitude for the null momenta $k_{[ij]}$. It would be interesting to see if there is a more general connection between tree-level amplitudes and the Newton polygons, or related structures, for soliton solutions. 

\subsection{X-solitons}
Continuing with increasing numbers of phases, we consider solutions with four  exponentials, 
$
	u=\sum_{i=1}^4 q_i E_i$, which leads to X-solitons with four soliton arms.
For example, if we choose
 \begin{align}
 	\lpangle q_1, q_3\rpangle=	\lpangle q_1, q_4\rpangle=	\lpangle q_2, q_3\rpangle=	\lpangle q_2, q_4\rpangle=0
\end{align}
or equivalently, 
\begin{align}
	f(\ell,m)=\frac{c_1 e^{\theta_1}+c_2 e^{\theta_2}}{d_3 e^{\theta_3}+d_4 e^{\theta_4}}~,
\end{align}
there are only two asymptotic values for $f$, i.e. zero and infinity. We consider a configuration where the regions defined by dominant exponentials in the numerator are not adjacent. That is, if we define $\rho_i=\text{Re}(\theta_i)+\log\|q_i\|$ then we have four regions $R_i=\{\rho_i>\rho_j\, \forall j\neq i\}$. In $R_1$ and $R_2$ the numerator exponentials dominate in the asymptotic region and $|f|\to \infty$ while in $R_3$ and $R_4$ the denominator dominates so that $|f|\to 0$.  We choose parameters such that two adjacent regions $R_i$ and $R_j$ also have $\lpangle q_i, q_j \rpangle=0$. Thus each soliton arm separates two regions where $|f|$ takes different asymptotic values i.e. alternating between zero and positive infinity. In the asymptotic regions, the soliton arms, $[ij]$, are again fixed by the conditions
\begin{align}
	\text{Re}\big[L_{[ij]}\big]=0~
\end{align}
  and so each arm is again characterised by a four-momentum $k^\mu_{[ij]}=\text{Re}[p^\mu_i-p^\mu_j]$ where $\theta_i=p_i\cdot x$ which thus satisfy the conservation law
\begin{align}
k^\mu_{[13]}+k^\mu_{[41]}+k^\mu_{[24]}+k^\mu_{[32]}=0~.
\end{align}
For the single-wall solitons and Y-solitons, the skeleton of the profile, that is ignoring its modulation, in the $\tilde{z}-w$ plane is simply translated at different values of $z$ and $\tilde{w}$. 
\begin{figure}
	\centering
		\centering
		\raisebox{-0.5\height}{\includegraphics[scale=0.3]{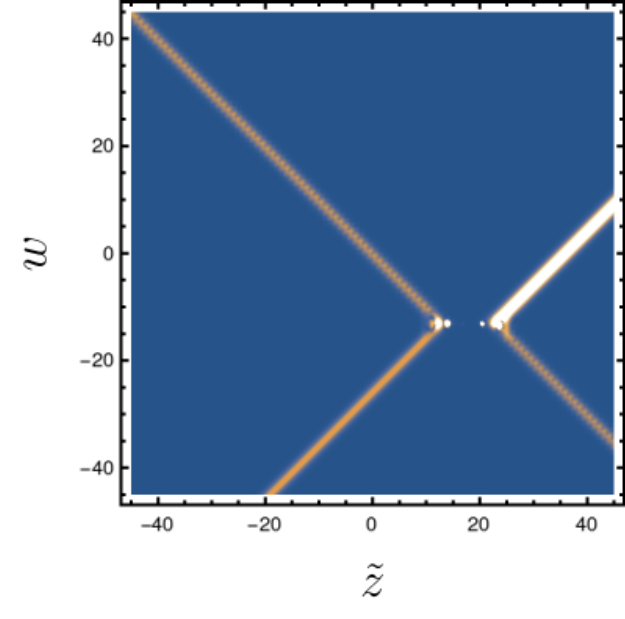}}~	\raisebox{-0.5\height}{\includegraphics[scale=0.3]{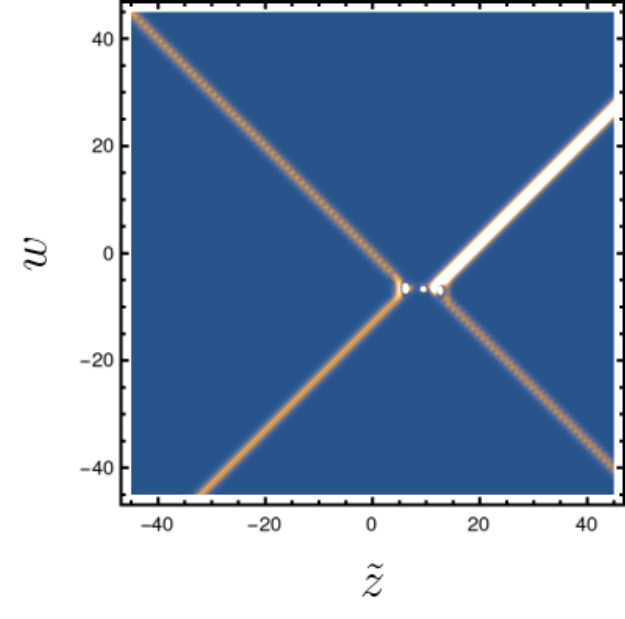}}~
		\raisebox{-0.5\height}{\includegraphics[scale=0.3]{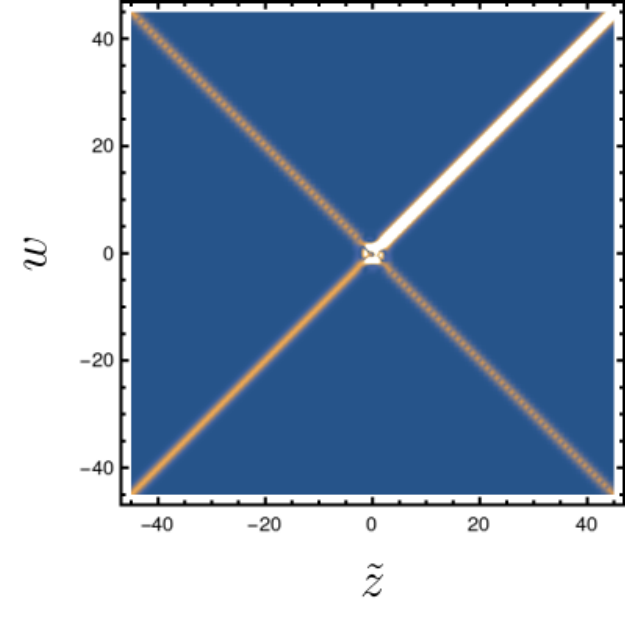}}~
		\raisebox{-0.5\height}{\includegraphics[scale=0.3]{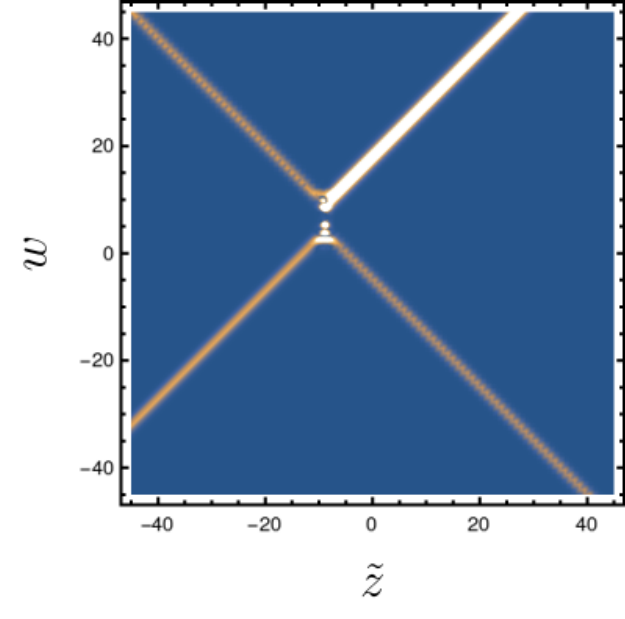}}~
		\raisebox{-0.5\height}{\includegraphics[scale=0.3]{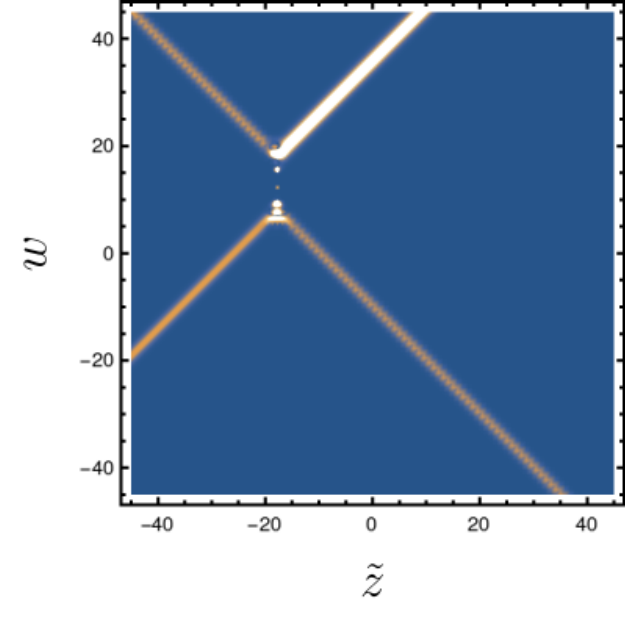}}
		\\
		\caption{ Density plots of $|\sigma|$ for the four-phase soliton with $\mu=i$, $a_{\ell,1}=1$, $a_{\ell,2}=-1$, $a_{\ell,3}=-i$, $a_{\ell,4}=i$, $a_{m,i}=e^{i\gamma_i}$ with $\gamma_i=(-\pi/\sqrt{2},\pi/\sqrt{5},\pi/2,11\pi/10)$, $c_1=c_2=d_3=d_4=1$. From left to right, plots are at  $z=\{-20,-10,0,10,20\}$ and $\tilde{w}=0$. }
	\label{fig:Xsolden}
\end{figure}
For the X-soliton this is not the case and instead when restricted to a specific plane the quartic vertex generically decomposes into two Y-solitons, see Fig. \ref{fig:Xsolden}. The quartic vertex corresponds to the space-time line defined by the intersection of four walls at $\rho_1=\rho_2=\rho_3=\rho_4$ which does not intersect with a generic two-plane. Vertices of the Y's occur along codimension-two surfaces in space-time defined by conditions $\rho_i=\rho_j=\rho_k$. The Y-vertices which comprise the X are joined by stems which correspond to a balance between terms in the numerator or denominator depending on which are dominant at a given $z$. The stems and soliton arms are dressed by colour factors which can modulate their appearance. While this bears some resemblance to the O-type interaction seen in the KP equation e.g. \cite{chakravarty2009soliton}, there is in fact no non-trivial scattering; to see that we will need to consider two-pole solutions.

\subsection{Soliton foams}
\label{sec:NPS}
For a general $n$-phase solution, describing up to $n$ distinct dominance regions with walls along their shared boundaries, we take $u$ to be a sum over $n$ exponentials
\begin{align}
	u=\sum_{i=1}^n q_i E_i =A E~~~\text{where}~~~E_i=\text{exp}(\theta_i), ~~~ \theta_i=p_i \cdot x
\end{align}
with $p_i =\lambda_i\tilde{\lambda}$ complex null momenta with $\lambda_{A,i}=(a_{\ell,i},a_{m,i})$, $\tilde{\lambda}_{A'}=(1,\mu)$. The $\tau$-function is then 
\begin{align}
	\tau =\sum_{i,j=1}^n G_{ij} \bar{E}_i E_j~, ~~~G=A^\dagger A
\end{align}
where $G$ is thus positive semi-definite. We again introduce the variables $\rho_i=\text{Re}(\theta_i)+\log\|q_i\|$ and define the regions of space-time $R_i=\{\rho_i>\rho_j,\, \forall j\neq i\}$.
We have seen previously that in each region $R_i$, the $\tau$-function is dominated by one diagonal term $e^{2\rho_i}$ which is large compared to the others. The soliton walls lie along the boundaries of these regions where Re$[L_{[ij]}]=\rho_i-\rho_j=0$ and  multiple walls can intersect on lower-dimensional surfaces, lines etc. Taking inspiration from the study of solitons in the KP equation e.g. \cite{kodama2011kp} we can map these elements of the solution onto a contour plot which approximates the positions where scalar densities $|\sigma|$ and $|\mathcal{L}|$ have peaks. 
For a specific solution characterised by the parameters $B=(A, p, \mu)$, we can approximate the logarithm of the $\tau$-function by 
\begin{align}
	\log \tau_B(x) = \log \sum_{ij} G_{ij} \bar{E}_i E_j \to s_B(x)=2 \max_i (\text{Re}(p_i)\cdot x +\log \|q_i\|)~.
\end{align}
The contour plot is the corner locus where the function $s_B(x)$ is not linear, that is to say the set of points where two, or more, different $\rho_i$'s are maximal and equal. The regions $R_i$ are convex polyhedra and the corner locus is the union of their shared boundaries. We can alternatively take an approach closer to that used in KP and define the contour as the corner locus of $s_B(\ell, m=m_0)$ at fixed $m=m_0$ which corresponds to taking slices of the four-dimensional contour on a null plane. This skeleton will be a tropical hypersurface and it would be interesting to study whether there is a useful combinatorial description of the contour plots for these one-pole, and indeed the general, multi-pole solutions.

 The tropical skeleton gives the set of walls each of which is further dressed by the colour data which can alter the soliton amplitude. We can rewrite \eqref{eq:sigma_def} as
 \begin{align}
 	\sigma=\kappa \frac{\langle \bar{W},W\rangle }{\|u\|^4}~, ~~~\text{with}~~~W_A=\frac{1}{2} \sum_{i,j=1}^n (\lambda_{A,i}-\lambda_{A,j}) \det(q_i, q_j) E_i E_j~.
 \end{align}
Near a particular wall $[ij]$, where the sum is dominated by two exponentials, we can write the scalar density as 
\begin{align}
\left.	\sigma\right|_{[ij]}= \frac{k^2_{[ij]} \sin^2 \chi_{ij} }{2(\cosh \text{Re}\, L_{[ij]}+\cos\chi_{ij} \cos \Phi_{ij})^2}
\end{align}
where $k_{[ij]}=\text{Re}[p_i-p_j]$, $\cos \chi_{ij}e^{i \phi_{ij}}=\lpangle q_i, q_j\rpangle/\|q_i\|\|q_j\|$, $L_{[ij]}=\theta_i-\theta_j+\log \tfrac{\|q_i\|}{\|q_j\|}$ and $\Phi_{ij}=\phi_{ij}-\text{Im}L_{[ij]}$. From this expression we can see that the wall has vanishing amplitude when $q_i\parallel q_j$, is modulated for generic orientations and is uniform for $q_i \perp q_j$ as we have seen in explicit examples. The soliton can also possess finite-action defects where $u$ has non-removable zeros. This condition corresponds to the vanishing of two holomorphic functions of two variables and so will generally be satisfied at isolated points.

\section{Two-pole scattering solutions}
We now turn to the case of two-pole solutions, which we refer to as two-soliton solutions. 
For this case the soliton matrix is given by \footnote{Here we enclose the indices on $u^{k}$ and related variables with brackets to avoid notational confusion with numerical exponents and as they can be usefully thought of as soliton labels.}
\begin{align}
	\Gamma=\begin{pmatrix} \frac{\lpangle u^{(1)}, u^{(1)}\rpangle}{\bar{\mu}_1-\mu_1} &  \frac{\lpangle u^{(1)}, u^{(2)}\rpangle}{\bar{\mu}_1-\mu_2}\\
		\frac{\lpangle u^{(2)}, u^{(1)}\rpangle}{\bar{\mu}_2-\mu_1} & \frac{\lpangle u^{(2)}, u^{(2)}\rpangle}{\bar{\mu}_2-\mu_2}
		\end{pmatrix}
\end{align}
where we assume $\mu_1\neq \mu_2$, $\mu_1 \neq \bar{\mu}_2$ and Im$\,\mu_i\neq 0$ for $i=1,2$.
We have for the $\tau$-function \eqref{eq:tau}
\begin{align}
	\tau=\kappa_1 \kappa_2 \text{det}\, \Gamma=\|u^{(1)}\|^2\|u^{(2)}\|^2\left(1-\gamma \frac{|\lpangle u^{(1)}, u^{(2)}\rpangle |^2}{\|u^{(1)}\|^2\|u^{(2)}\|^2}\right)
\end{align}
using the notation $\kappa_i=\mu_i-\bar{\mu}_i$ and where we have introduced the combinations
\begin{align} 
	\gamma=\frac{(\bar{\mu}_1-\mu_1)(\bar{\mu}_2-\mu_2)}{(\bar{\mu}_1-\mu_2)(\bar{\mu}_2-\mu_1)}=1-|r|^2\leq1~,~~~\text{with}~~~r=\frac{\mu_1-\mu_2}{\mu_1-\bar{\mu}_2}~.
\end{align}
Using the Lagrange identity $\|u^{(1)}\|^2\|u^{(2)}\|^2=|\lpangle u^{(1)}, u^{(2)}\rpangle|^2+|\det(u^{(1)}, u^{(2)})|^2$ we can re-write this as 
\begin{align}
\tau=|\det(u^{(1)}, u^{(2)})|^2+|r|^2 |\lpangle u^{(1)}, u^{(2)}\rpangle|^2
\end{align}
which is manifestly positive for $u^{(k)} \neq 0$ as expected. 

This can be used to compute scalar quantities characterising the solution. Using the freedom to rescale each $u^{(k)}$ we can again introduce the inhomogeneous coordinates 
\begin{align}
	u^{(1)}=\begin{pmatrix} f^{(1)}(\ell_1, m_1) \\ 1 \end{pmatrix}~, ~~~u^{(2)}=\begin{pmatrix} f^{(2)}(\ell_2, m_2) \\ 1 \end{pmatrix}
\end{align}
and in terms of the holomorphic functions $f^{(1)}$ and $f^{(2)}$ we have for the scalar density $\sigma$ the relatively compact expression:
\begin{align}
	\sigma&=\Box \log \tau=\frac{\kappa_1 \kappa_2}{\tau^2}\left|r\right|^2\Big(\kappa_1 (\Gamma^{{1}1})^2\langle\bar{f}^{(2)},f^{(2)}\rangle+ \kappa_2 (\Gamma^{{2}2})^2\langle \bar{f}^{(1)},f^{(1)}\rangle
	\nn\\
	&-2\text{Re}\big[(\mu_1-\bar{\mu}_2)(\Gamma^{{2}1})^2\langle \bar{f}^{(1)},{f}^{(2)}\rangle \big]
	+2\text{Re}\left[\frac{(f^{(1)}-f^{(2)})^2}{\mu_1-\mu_2}\langle\bar{f}^{(1)},\bar{f}^{(2)}\rangle \right]\Big)
\end{align}
where for the spinors $f^a_A=(\partial_\ell f^a, \partial_m f^a)$, $\bar{f}^a_A=(\partial_{\bar{\ell}} \bar{f}^a, \partial_{\bar{m}} \bar{f}^a)$ we have used the brackets
\begin{align}
& \langle \bar{f}^a, f^b\rangle =\partial_{\bar{\ell}_a}\bar{f}^a \partial_{m_b}f^b-\partial_{\bar{m}_a} \bar{f}^a \partial_{\ell_b}f^b\nn\\
& \langle \bar{f}^a, \bar{f}^b\rangle =\partial_{\bar{\ell}_a}\bar{f}^a \partial_{\bar{m}_b}\bar{f}^b-\partial_{\bar{m}_a} \bar{f}^a \partial_{\bar{\ell}_b}\bar{f}^b
\end{align}
In this case we do not have a compact form for the Lagrangian density. However, we have numerically checked that for a range of solutions and parameters the relation 
\begin{align}
	\mathcal{L}=-\Box^2 \log \tau
\end{align}
holds in the two-soliton sector and, as we will see below, this is a general result which holds for arbitrary numbers of poles.

\subsection{Extended-wave scattering}
\begin{figure}
	\centering
	\raisebox{-0.5\height}{\includegraphics[scale=0.35]{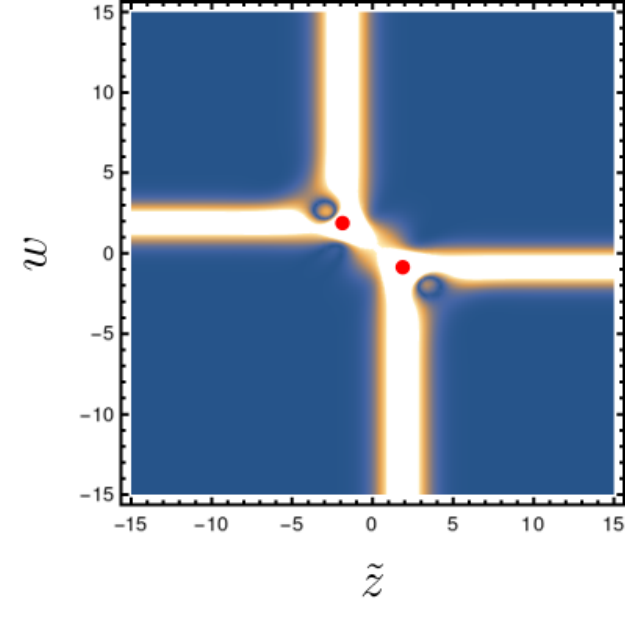}}~
	\raisebox{-0.5\height}{\includegraphics[scale=0.35]{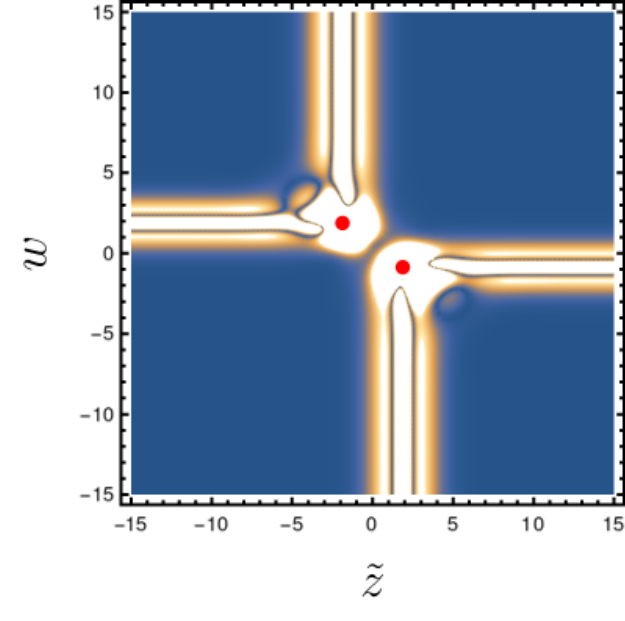}}~
	\raisebox{-0.5\height}{\includegraphics[scale=0.35]{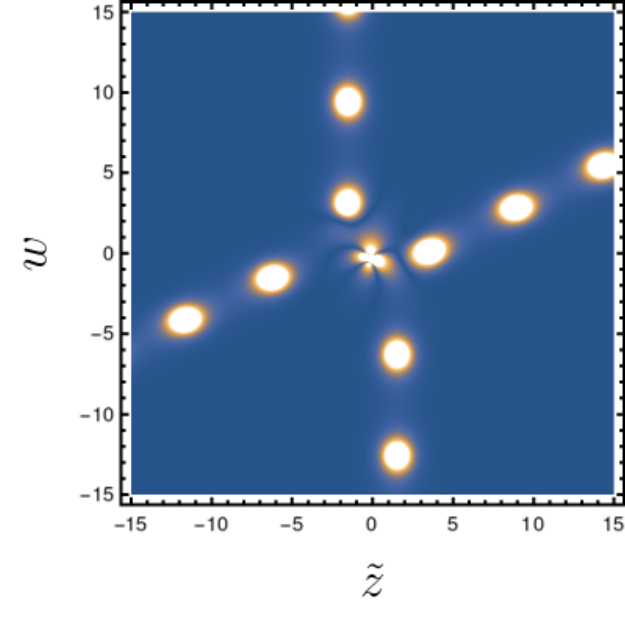}}
		\raisebox{-0.5\height}{\includegraphics[scale=0.35]{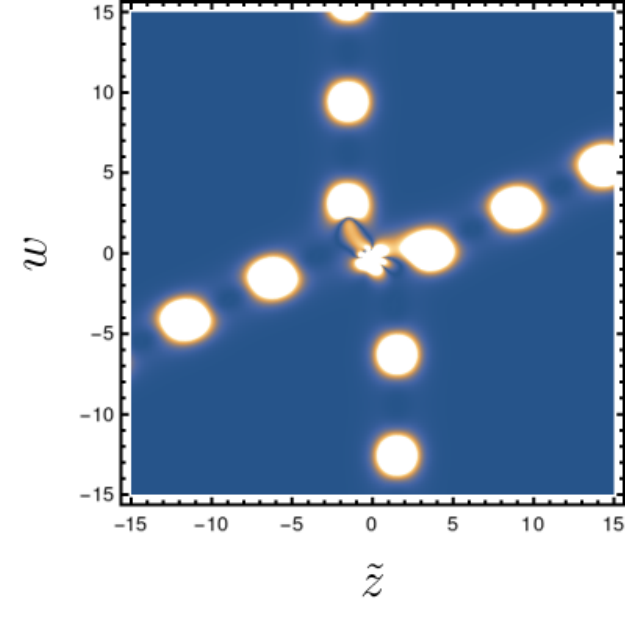}}

	\caption{Left: Density plot of $|\sigma|$ for the two-soliton extended-wave solution with $\mu_1=i$, $\mu_2=1.36 i$, $m={z}=\tilde{w}=0$, $a_{\ell}^{(1)}=1$, $a_{\ell}^{(2)}=i$,  $c^{(1)}=1$, $c^{(2)}=-2$. Red dots are locations of vertices at $\phi^{(1)}=\phi^{(2)}=\pm \log |r|$.  Centre Left: Lagrangian density for two-soliton extended wave. Centre Right: Density plot of $|\sigma|$ for the two-soliton two-phase solution with $\mu_1=i$, $\mu_2=1.2 i$, $m={z}=\tilde{w}=0$, $a_{\ell,1}^{(1)}=1$, $a_{\ell,1}^{(2)}=e^{i\pi/3}$, $a_{\ell,2}^{(1)}= a_{\ell,2}^{(2)}=0$, $q_1^{(1)}=(1,0)$, $q_2^{(1)}=(0,1)$, $q_1^{(2)}=(\cos \pi/16, \sin \pi/16)$, $q_2^{(2)}=(-\sin \pi/16, \cos \pi/16)$. Right:  Lagrangian density for two-soliton two-phase solution.  }
	\label{fig:TwopoleEWsolden}
\end{figure}
The two-soliton sector can be used to compute the phase shift acquired by one soliton as it passes through another. 
The simplest case is to consider the solution with two extended waves
\begin{align}
u^{(1)}=\begin{pmatrix} c^{(1)} e^{\theta^{(1)}}\\ 1\end{pmatrix}~,~~~u^{(2)}=\begin{pmatrix} c^{(2)} e^{\theta^{(2)}}\\ 1\end{pmatrix}
\end{align}
where $\theta^{(i)}=p^{(i)} \cdot x$ and $p^{(i)}$ are  complex null momenta
but we now have $p^{(i)}=\lambda^{(i)} \tilde{\lambda}^{(i)}$ with $\lambda^{(i)}=(a^{(i)}_{\ell}, a^{(i)}_{m})$ and $\tilde{\lambda}^{(i)}=(1,\mu_i)$
with different, complex $\mu_i$'s for the two solitons. 
We then consider the change in the phase of the first soliton as the solution interpolates between the asymptotic regions of the second:
\begin{align}
	&\tau\simeq |r|^2(|r|^{-2}e^{2\phi^{(1)}}+1)~~~\text{as}~~~\text{Re}\,\theta^{(2)}\to -\infty\nn \\
	&\tau\simeq  e^{2 \phi^{(2)} }(|r|^2e^{2\phi^{(1)}}+1)~~~\text{as}~~~\text{Re}\, \theta^{(2)}\to +\infty
\end{align}
where $\phi^{(i)}=\text{Re}\,\theta^{(i)}+\log |c^{(i)}|$ is the same phase appearing in the one-pole solution \eqref{eq:ext_wv_1}. The profile of the first soliton is thus shifted from $\phi^{(1)}=\log |r|$ in one asymptotic region to $\phi^{(1)}=-\log |r|$ in the other. By symmetry, the same result is found for the shift of the second soliton due to interaction with the first  and so each wall is phase shifted by 
\begin{align}
	\label{eq:sph}
	\Delta =-\log \left|\frac{\mu_1-\mu_2}{\mu_1-\bar{\mu}_2}\right|^2~.
\end{align}
This is the same expression as found in the 2+1-dimensional integrable chiral model \cite{Leese:1989hj} for the scattering of extended waves and in \cite{Li:2025dlc} for ASDYM, see also \cite{Hamanaka:2022ole}; the same form is also found in sine-Gordon scattering. The phase shift is invariant under Lorentz transformations as can be seen by checking invariance under SL$(2, \mathbbm{R})$ or by noting that it can be written in terms of Lorentz invariant spinor brackets
\begin{align}
\Delta =-\log \left| \frac{[\tilde{\lambda}^{(1)},\tilde{\lambda}^{(2)} ]}{[\tilde{\lambda}^{(1)},\bar{\tilde{\lambda}}^{(2)} ]}\right|^2~.
\end{align}
The phase shift also has a nice interpretation in terms of the profile of the solution as explained in \cite{Hamanaka:2022ole, Li:2025dlc} where the two walls form two Y-type solitons with vertices at $\phi^{(1)}=\phi^{(2)}=\pm \log |r|$, joined along a stem with the phase shift being the length of the stem, see Fig. \ref{fig:TwopoleEWsolden} Left and centre left panels.

\subsection{Two-phase wall scattering}

Wall scattering is strongly dependent on the colour orientation of the individual walls and to find a colour-dressed scattering phase it is useful to consider the two-phase wall solutions. That is, we take 
\begin{align}
	u^{(i)}=q_1^{(i)}E_1^{(i)}+q_2^{(i)}E_2^{(i)}
\end{align}
where $E^{(j)}_i=\text{exp}(\theta^{(j)}_i)$ is the $i$-th exponential of the $j$-th soliton. 
We follow the motion of the first soliton in the background of the second. We take the incoming asymptotic region to be $R^{(2)}_2$ defined by Re$\,\theta^{(2)}_2\to \infty $ and the outgoing region to be $R^{(2)}_1$ with Re$\,\theta^{(2)}_1\to \infty $. In $R^{(2)}_2$ we have for the $\tau$-function
\begin{align}
	\tau\simeq |E_2^{(2)}|^2\|q_2^{(2)}\|^2 \left(\bar{u}^{(1)}{}^T\big[1-\gamma P_2^{(2)}\big] u^{(1)}\right)~,~~~\text{with}~~~P_2^{(2)}=\frac{q_2^{(2)}\otimes q_2^{(2)}{}^\dagger}{\|q_2^{(2)}\|^2}
\end{align}
where we see that the dependence of $\tau$ on the first soliton has essentially the same structure as in the one-pole case, \eqref{eq:tau_two_phase}, but with the scalar product $\lpangle \cdot, \cdot \rpangle$ replaced by $\lpangle \cdot, \cdot \rpangle^{(2)}\equiv\lpangle \cdot, \big[1-\gamma P_2^{(2)}\big]\cdot \rpangle$. Consequently we can take the expressions of Sec. \ref{sec:two_phase_wall} but use the modified scalar product. In particular, 
\begin{align}
	\label{eq:tau_bk_two_phase}
\tau\propto & \, \bar{u}^{(1)}{}^T\big[1-\gamma P_2^{(2)}\big] u^{(1)}
\propto  \cosh \text{Re}[L']+\cos \chi' \cos \Phi'
\end{align}
with 
\begin{align}
L'=\theta_1^{(1)}-\theta_2^{(1)}+\log \frac{\|q_1^{(1)}\|^{(2)}}{\|q_2^{(1)}\|^{(2)}}
~,~~~
	\cos \chi' 
	e^{i \phi'}
	=\frac{
		\lpangle q_1^{(1)},q_2^{(1)} \rpangle^{(2)}}{\|q_1^{(1)}\|^{(2)}\|q_2^{(1)}\|^{(2)}}~,
\end{align}
and $\Phi'=\phi'-\text{Im} (L')$.
Thus in region $R^{(2)}_2$ the first soliton is centred on the surface
\begin{align}
	\text{Re}[\theta_1^{(1)}-\theta_2^{(1)}]=\log \frac{\|q_2^{(1)}\|^{(2)}}{\|q_1^{(1)}\|^{(2)}}=\log \frac{\|q_2^{(1)}\|}{\|q_1^{(1)}\|}+\frac{1}{2} \log \frac{1-\gamma \cos^2 \chi^{(1)(2)}_{22}}{1-\gamma \cos^2\chi^{(1)(2)}_{12}}
\end{align}
where we use the notation $\lpangle q_i^{(\alpha)}, q_j^{(\beta)}\rpangle /( \|q_i^{(\alpha)}\|\|q_j^{(\beta)}\|)=\cos \chi_{ij}^{(\alpha)(\beta)}e^{i\phi_{ij}^{(\alpha)(\beta)}}$. 

Repeating for the region $R_1^{(2)}$ we see that the phase shift of the first soliton is given by 
\begin{align}
	\label{eq:cdph}
	\Delta^{(1)} =\frac{1}{2} \log \frac{(1-\gamma \cos^2\chi^{(1)(2)}_{12})(1-\gamma \cos^2\chi^{(1)(2)}_{21})}{(1-\gamma \cos^2 \chi^{(1)(2)}_{11})(1-\gamma \cos^2 \chi^{(1)(2)}_{22})}
\end{align}
which by symmetry is the same as the phase shift of the second soliton, $\Delta^{(2)}$, as it moves from $R_2^{(1)}$ to $R_1^{(1)}$. This phase shift shares the Lorentz invariance of the one-phase case \eqref{eq:sph} as $\gamma$ is invariant under SL$(2, \mathbbm{R})$ and \eqref{eq:cdph} generalises that result to a colour-dressed phase which accounts for walls with non-aligned orientations. If we choose $q_1^{(k)}\perp  q_2^{(k)}$ for $k=1,2$ and $\chi_{11}^{(1)(2)}=\chi_{22}^{(1)(2)}=\alpha$, so that the colour frame of one wall is rotated by $\alpha$ compared to the other, we get a simple form of the phase shift
\begin{align}
	\Delta = \log \frac{1-\gamma \sin^2\alpha}{1-\gamma \cos^2\alpha}
\end{align}
 which gives the aligned result $\Delta=-\log |r|^2$ for $\alpha=0$, but a vanishing result for $\alpha=\pi/4$ and the opposite result $\Delta=+\log |r|^2$ for $\alpha=\pi/2$. Thus we see that the colour orientation can allow for the walls to pass through each other with no shift or reverse the shift. 
 
We can plot the scalar density $\sigma$, Fig. \ref{fig:TwopoleEWsolden} centre right panel, and we see that it is similar to the one-phase case though now with modulations along the soliton arms as was found in the two-phase one-pole solutions Fig. \ref{fig:2phase1Ew}. This modulation is due to the phases $\Phi'$ when the colour factors $\chi'$ are not equal to $\pm \pi/2$. These phases and colour factors change when passing from one asymptotic region to another and so in addition to a phase shift there can be a change in the amplitude of the soliton arms. 

This method can be extended to solutions with higher numbers of phases. For a given pole, each phase defines an asymptotic region and in that region one can compute the soliton arms of the other pole. They will be as in the one-pole solution but with a modified scalar product which produces a phase shift and a modified amplitude. Following the soliton arm to another asymptotic region produces a phase shift and change in amplitude which follow from the above expressions. Crossing through multiple asymptotic regions produces phase shifts which add linearly. 

\subsection{Lump-wall scattering}
\begin{figure}
	\centering
	\raisebox{-0.5\height}{\includegraphics[scale=0.3]{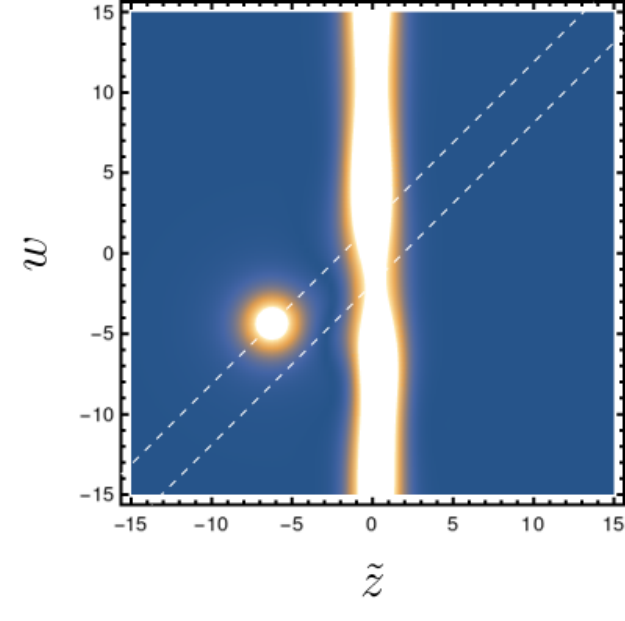}}~	\raisebox{-0.5\height}{\includegraphics[scale=0.3]{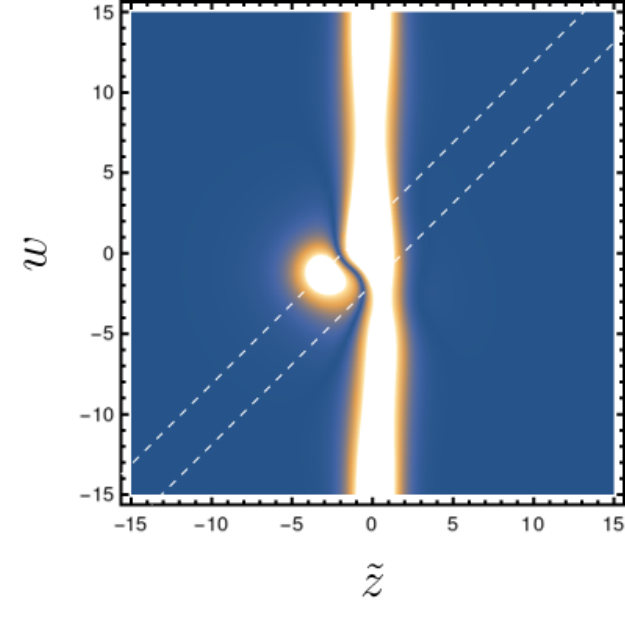}}~
	\raisebox{-0.5\height}{\includegraphics[scale=0.3]{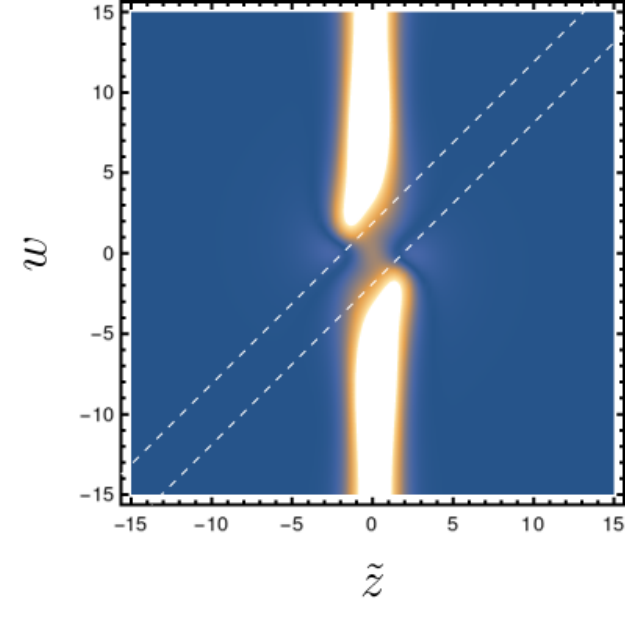}}~
	\raisebox{-0.5\height}{\includegraphics[scale=0.3]{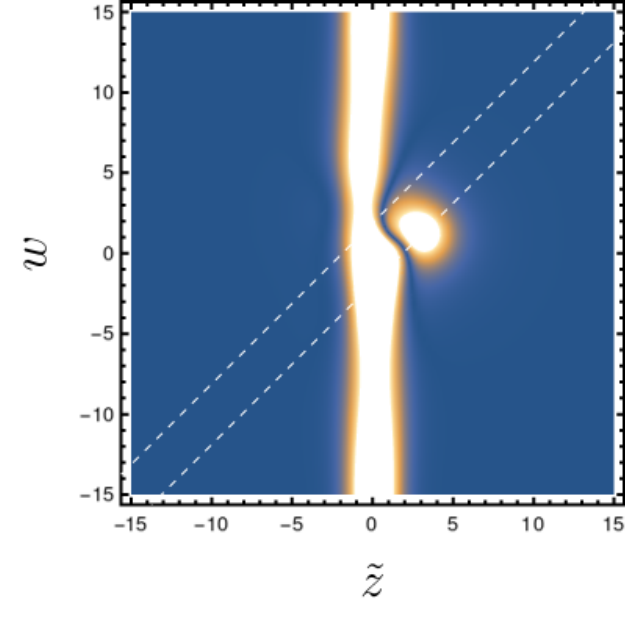}}~
	\raisebox{-0.5\height}{\includegraphics[scale=0.3]{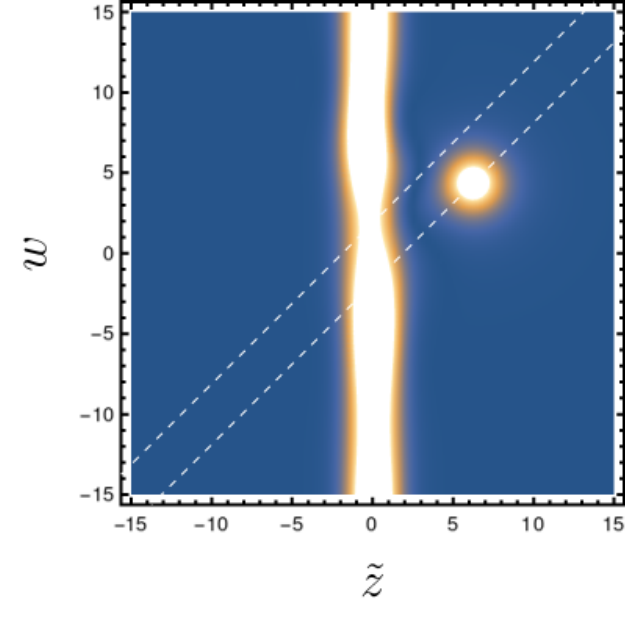}}
	\caption{ Density plots of $|\sigma|$ for the scattering of a lump and a wall with $\mu_1=i$, $\mu_2=i/2$, $f=e^{i\pi/4}\ell_1-\tfrac{3i}{2} m_1$,  $a^{(2)}_{\ell,1}=1$, $a^{(2)}_{\ell,2}=0$, $a^{(2)}_{m,1}=\tfrac{23i}{5}$, $a^{(2)}_{m,2}=0$, $c^{(1)}_1=1$, $d^{(1)}_1=0$, $c^{(1)}_2=0$, $d^{(1)}_2=1$, $c^{(2)}_1=c_2^{(2)}=1$, $d^{(2)}_1=-d^{(2)}_2=-1/3$. From left to right, plots are at  $z=\{-5,-2,0,2,5\}$ and $\tilde{w}=0$. Dashed lines show trajectory of lump before and after scattering. }
	\label{fig:LW}
\end{figure}
It is also possible to consider the scattering of a lump soliton from a wall. We take 
\begin{align}
	u^{(1)}= q^{(1)}_1 f +q_2^{(1)} ~,~~~
	u^{(2)}=q_1^{(2)} e^{\theta_1^{(2)}}+q_2^{(2)} e^{\theta_2^{(2)}}
\end{align}
where $f$ is a polynomial or rational function and $\theta^{(2)}_i=p_i^{(2)}\cdot x$ with $p_i^{(2)}$ complex null momenta. Up to a M\"obius transformation of $f$, we may take $\|q_1^{(1)}\|=\|q_2^{(1)}\|$ and $\lpangle q_1^{(1)}, q_2^{(1)}\rpangle=0$ without loss of generality. We again compute the $\tau$-function in the two asymptotic regions defined by the wall soliton $R^{(2)}_1$ where Re$\,\theta_1^{(2)}\to \infty$ and $R^{(2)}_2$ where Re$\,\theta_2^{(2)}\to \infty$. In $R_2^{(2)}$ we have that, similar to \eqref{eq:tau_bk_two_phase},
\begin{align}
	\tau&\simeq \,|E_2^{(2)}|^2\|q_2^{(2)}\|^2 (\bar{u}^{(1)}{}^T\big[1-\gamma P_2^{(2)}\big] u^{(1)})
	\propto |f-f^{(2)}_2|^2+ w_{2}^{(2)}
\end{align}
with 
\begin{align}
	f^{(2)}_2
	=\gamma  \frac{\cos \chi_{12}^{(1)(2)} \cos \chi_{22}^{(1)(2)} e^{i (\phi_{12}^{(1)(2)}-\phi_{22}^{(1)(2)})}}{1-\gamma \cos^2 \chi_{12}^{(1)(2)}}
~,~~~\text{and}~~~
	w_{2}^{(2)}=
	\frac{|r|^2}{(1-\gamma \cos^2 \chi_{12}^{(1)(2)})^2}~.
\end{align}
We can see that the centre of the lump is shifted from $f=0$ by the presence of the extended wave to $f^{(2)}_2$ which depends on the wall's orientation in colour space as does the shape of the lump which is determined by $w_2^{(2)}$. In region $R^{(2)}_1$ we have the same effect with the corresponding shift $f^{(2)}_1$ and shape parameter $w_1^{(2)}$. Upon crossing from region $R^{(2)}_2$ to $R^{(2)}_1$, i.e. crossing the wall soliton, the  lump is displaced by 
\begin{align}
	\Delta f=f^{(2)}_1-f^{(2)}_2
\end{align}
and its shape/amplitude changes, see Fig. \ref{fig:LW}. 

\subsection{Lump-lump scattering}
Unlike the wall soliton, a lump does not divide space-time into different asymptotic regions with different colour backgrounds. Given a lump solution, there is no way to define the notion of moving a second lump from one asymptotic region to another.  Thus, as in the case of the $2+1$ dimensional integrable chiral model, two lumps corresponding to distinct poles do not scatter. However, for specific kinematic configurations there can in fact be non-trivial scattering \cite{Sutcliffe:1992dj} and this was shown by Ward \cite{Ward:1995he} to arise from double-pole solutions. The construction of these solutions starts from an ansatz generalising the $N=1$ single-pole case of \eqref{eq:psi_int} but now with an additional double pole
\begin{align}
	\label{eq:psi_dp}
\psi(\zeta)=\mathbbm{1}+\frac{M(x)}{\zeta-\mu}+\frac{\tilde{M}(x)}{(\zeta-\mu)^2}~.
\end{align}
The generalised unitarity condition implies that this solution factorises
\begin{align}
	\label{eq:psi_fact}
\psi=(\mathbbm{1}+\frac{\kappa}{\zeta-\mu}P_2)(\mathbbm{1}+\frac{\kappa}{\zeta-\mu}P_1)
\end{align}
where $P_1$ and $P_2$ are projections onto two one-dimensional subspaces. Ward further showed that such a solution could be found by considering the $\epsilon \to 0$ limit of a two-pole solution where $\mu_1=\mu+\epsilon$, $\mu_2=\mu-\epsilon$. We can consider a minimal generalisation of Ward's construction to the current context by taking the polarisation vectors
\begin{align}
	u^{(1)}=\begin{pmatrix}f^{(1)}(a_\ell \ell_1+a_m m_1)\\ 1\end{pmatrix}~,~~~u^{(2)}=\begin{pmatrix}f^{(2)}(a_\ell \ell_2+ a_m m_2)\\ 1\end{pmatrix}
\end{align}
and further assuming that $f^{(1)}(x)=f(x)+\epsilon h(x)$, $f^{(2)}(x)=f(x)-\epsilon h(x)$ with $f$ and $h$ rational functions. One can show that $\psi$ is finite in the limit and takes the factorised form \eqref{eq:psi_fact} with 
\begin{align}
	\label{dp_proj}
	P_1&=\frac{\xi_1 \otimes \xi_1^\dagger}{\|\xi_1\|^2}~,~~~ \xi^\dagger_1=(f(\theta), 1)~\nn\\
	P_2&=\frac{\xi_2 \otimes \xi^\dagger_2}{\|\xi_2\|^2}~,~~~ \xi^\dagger_2=(1+|f(\theta)|^2)( f(\theta), 1)+\kappa g(\theta,t) (1, -\bar{f}(\bar{\theta}))
\end{align}
where $g(\theta,t)=t f'(\theta)+h(\theta)$ with $\theta=a_\ell \ell+a_m m$,  $t=a_\ell w+ a_m z$. The $\tau$-function can be computed for the two-pole solution and in the limit gives
\begin{align}
	\label{eq:taudp}
	\tau_\epsilon=-\frac{4\epsilon^2}{\kappa^2} \big[(1+|f|^2)^2-\kappa^2 |g|^2\big]+\mathcal{O}(\epsilon^4)~.
\end{align}
It is particularly interesting to note that this $\tau$-function can still be used to compute the Lagrangian density and we have numerically checked that 
\begin{align}
	\mathcal{L}=-\Box^2 \log \tau_{\text{dp}}~, ~~~\text{with}~~~\tau_{\text{dp}}=\lim_{\epsilon\to 0}\tau_\epsilon/\epsilon^2
\end{align}
continues to hold. Thus we have an example where our definition of the $\tau$-function holds even for double-pole solutions. 

\begin{figure}
	\centering
	\raisebox{-0.5\height}{\includegraphics[scale=0.3]{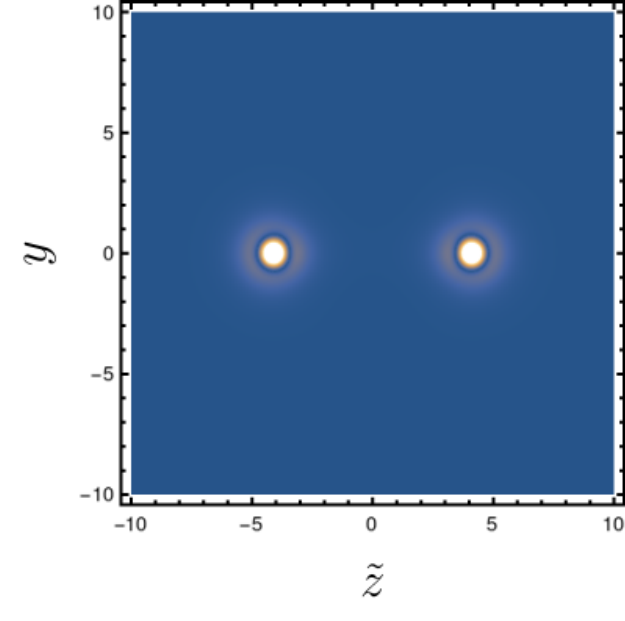}}~	\raisebox{-0.5\height}{\includegraphics[scale=0.3]{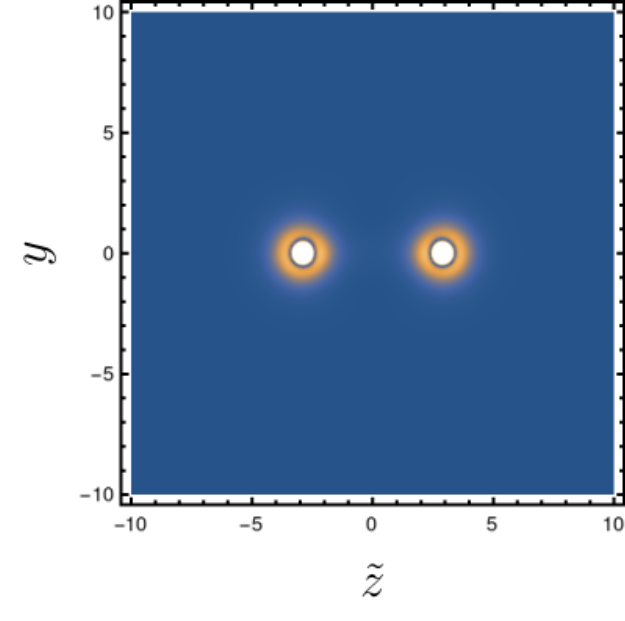}}~
	\raisebox{-0.5\height}{\includegraphics[scale=0.3]{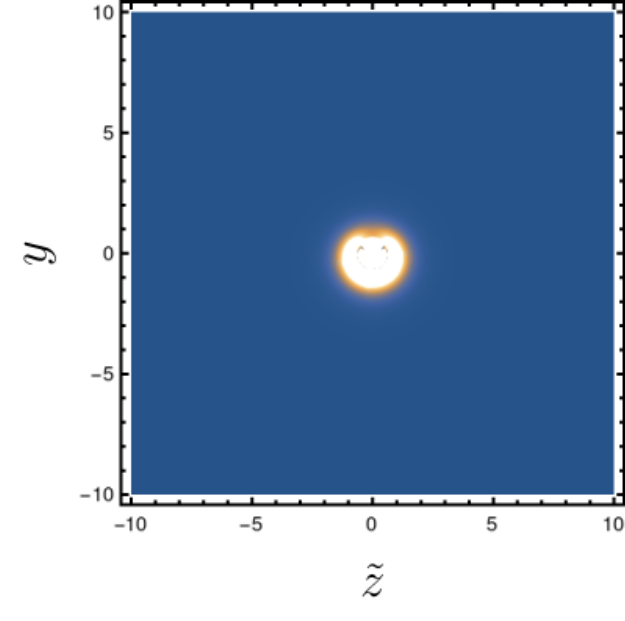}}~
	\raisebox{-0.5\height}{\includegraphics[scale=0.3]{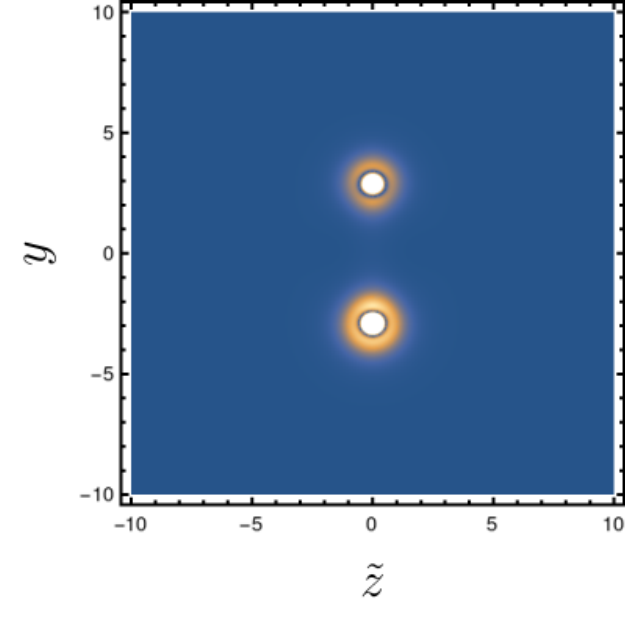}}~
	\raisebox{-0.5\height}{\includegraphics[scale=0.3]{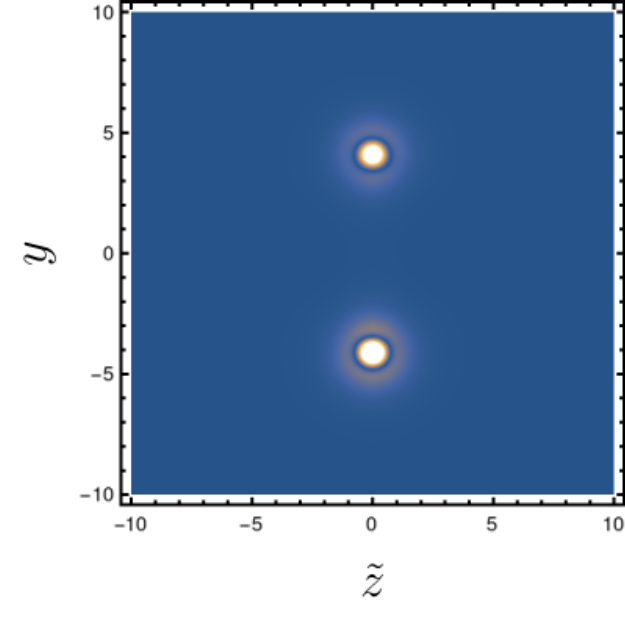}}\\
	\raisebox{-0.5\height}{\includegraphics[scale=0.3]{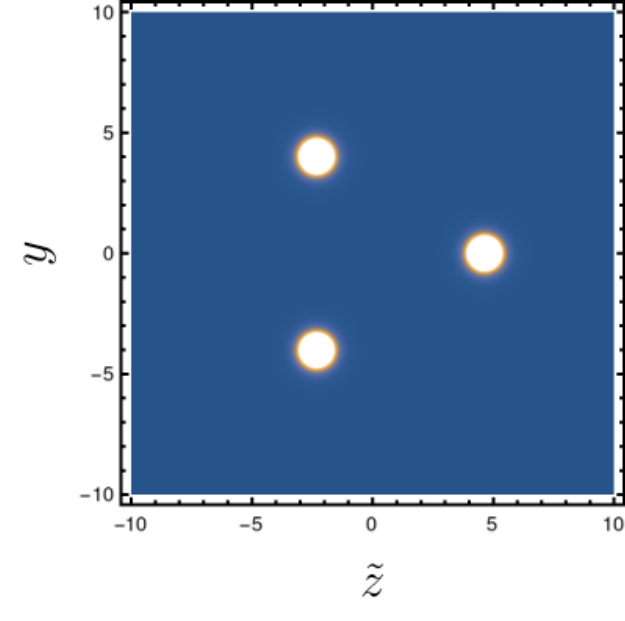}}~	\raisebox{-0.5\height}{\includegraphics[scale=0.3]{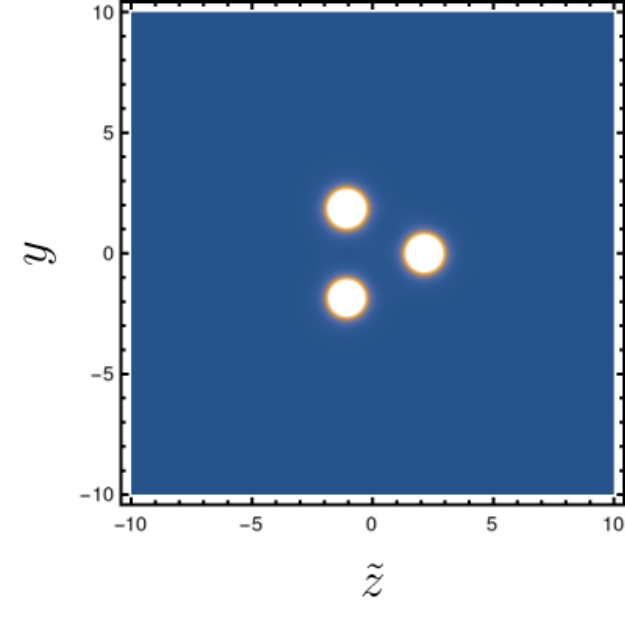}}~
	\raisebox{-0.5\height}{\includegraphics[scale=0.3]{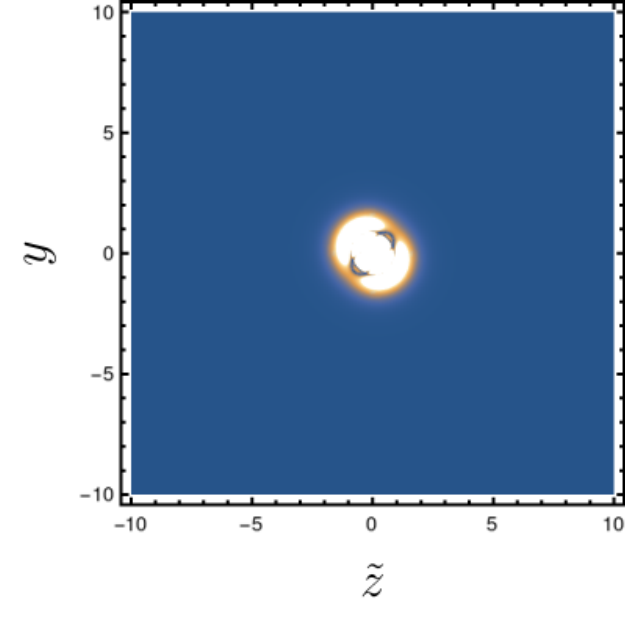}}~
	\raisebox{-0.5\height}{\includegraphics[scale=0.3]{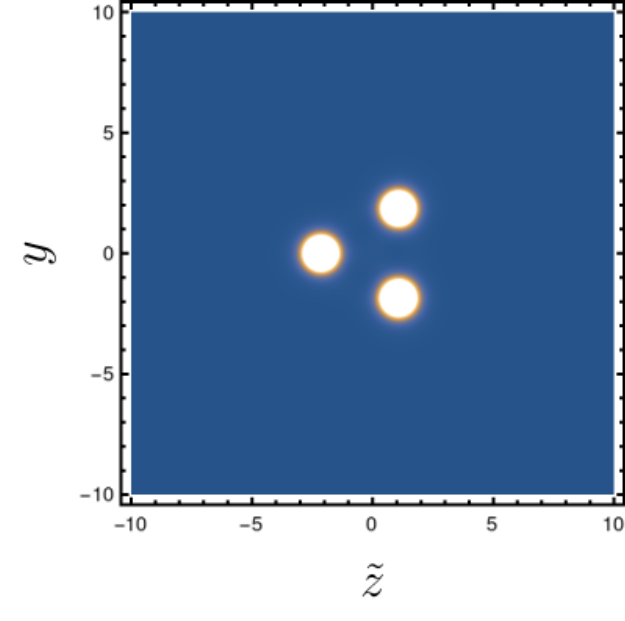}}~
	\raisebox{-0.5\height}{\includegraphics[scale=0.3]{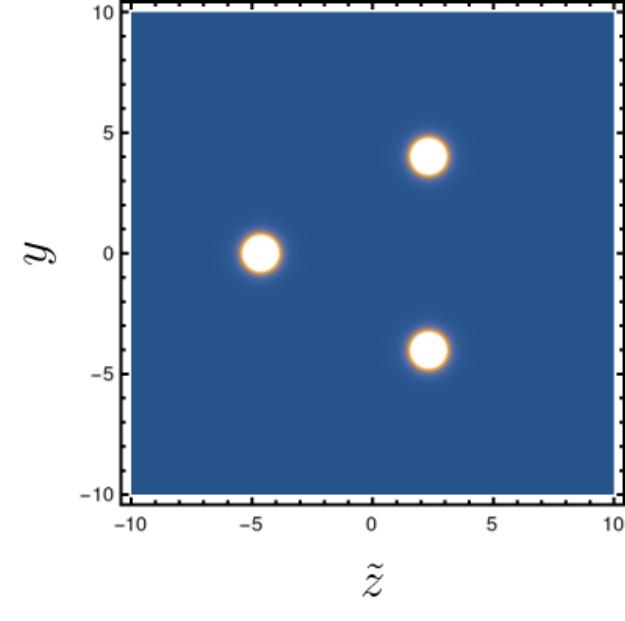}}
	\caption{Top: Density plots of $|\mathcal{L}(\tau)|$ for the lump-lump soliton with $\mu=i$, $f(x)=x$, $h(x)=x^2$, $a_{\ell}=1$, $a_{m}=i$. From left to right, plots are at  $w=\{-20,-10,0,10,20\}$, ${z}=0$ and $y=\tilde{w}+w$. Bottom: Same but for $h(x)=x^3$ and $w=\{-100,-10,0,10,100\}$~.}
	\label{fig:LLsol}
\end{figure}

The solutions can be seen to describe multi-lump scattering with constrained kinematics. Considering \eqref{eq:taudp}, we see that the locations of the lumps are determined by the condition $g=0$. For example, taking $f(\theta)=\theta$, $h(\theta)=\theta^2$ and so $g=t+ \theta^2$ we see that for fixed $t$ there are two solutions $\pm \sqrt{-t}$. Choosing a coordinate plane with $t$ real, we see that the two lumps approach head-on for $t<0$, collide at $t=0$ and scatter at right angles for $t>0$. Similarly, cubic $h(\theta)$ produces three lumps which rotate through $\pi/3$, see Fig. \ref{fig:LLsol}.

\section{Multi-pole solutions}
Finally we consider the case of general multi-pole solutions. Here we will not consider specific solutions but rather show that the same definition of the $\tau$-function holds for an arbitrary number of poles and so is a potential for interesting gauge-invariant scalar quantities for the entire class of solutions. The first step is to show that $\tau$ has a factorised structure following from the known properties of dressing solutions. 

The general $N$-pole solution has a useful rewriting as a product \cite{zakharov1978relativistically, zakharov1979integration, uhlenbeck1989harmonic}
\begin{align}
	\label{eq:psi_prod}
\psi^{(N)}(\zeta;\mu_1, \dots, \mu_N)=g^{(N)}(\zeta)\dots g^{(1)}(\zeta)~,~~~g^{(k)}(\zeta)=\mathbbm{1}+\frac{\kappa_k}{\zeta-\mu_k} P_{\xi^k}~,~~~ P_{\xi^k}=\frac{\xi^k\otimes \xi^k{}^\dagger}{\|\xi^k\|^2}
\end{align}
where the dressing vectors $\xi^k$ are defined iteratively by 
\begin{align}
	\xi^k=\psi^{(k-1)}(\bar{\mu}_k)\bar{u}^k~,~~~\text{with}~~~\psi^{(0)}=\mathbbm{1}.
\end{align}
Each dressing factor corresponds to a one-pole application of the solution generating procedure of Sec. \ref{sec:ZS} in the background of the previous dressings. Consequently the dressing vectors are covariantly anti-holomorphic $D^{(k)}_{\ell} \xi^k=	D^{(k)}_m \xi^k=0$ with
\begin{align}
 {D}_{\ell}^{(k)}=\frac{1}{\kappa_k}(\partial_{w}-\bar{\mu}_k \partial_{\tilde{z}}+A_{w}^{(k-1)}),~
	{D}_{m}^{(k)}=\frac{1}{\kappa_k}(\partial_{z}-\bar{\mu}_k \partial_{\tilde{w}}+A_{z}^{(k-1)})
\end{align}
where  $A_{a}^{(k-1)}$ is the potential determined from $\psi^{({k-1})}$ and $\partial_{\ell_k}\bar{u}^k=\partial_{m_k}\bar{u}^k=0$. %
In \eqref{eq:psi_prod} the order of $\mu_k$'s can be permuted, though different orderings require different sets of dressing factors \cite{terng2000backlund, Dai:2004fj}.  The product formula can be used to show the factorisation structure, see App. \ref{app:detfact},
\begin{align}
	\tau^{(N)}=\tau^{(N-1)} \|\xi^N\|^2=\prod_{k=1}^N  \|\xi^k\|^2~.
\end{align}
As a consequence, the scalar $\sigma$ is a sum of terms
\begin{align}
	\sigma^{(N)}=\Box \log \tau^{(N)}=\sum_{k=1}^N \sigma_k~,~~~\text{with}~~~\sigma_k=\Box \log \|\xi^k\|^2
\end{align}
where each term is given in terms of covariant derivatives of projection operators
\begin{align}
	\label{eq:sigmak}
\sigma_k&=\kappa_k \text{Tr}(D_{\bar{\ell}}^{(k)} P_{\xi^k}D_{m}^{(k)} P_{\xi^k}-D^{(k)}_\ell P_{\xi^k}D^{(k)}_{\bar{m}} P_{\xi^k})~,~~~\text{with}~~~D^{(k)}_a P_{\xi^k}= \partial_a P_{\xi^k}+[A^{(k-1)}_a,P_{\xi^k}]~.
\end{align}

\subsection{WZW$_4$ action}
We can show that $\sigma_k$ has an interesting relation to the WZW$_4$ Lagrangian \cite{Pohlmeyer:1979ya, Donaldson:1985zz, Nair:1990aa, Losev:1995cr} which has been recently used as a density for extended wave solitons in ASDYM \cite{Hamanaka:2022ole, Li:2025dlc}. The WZW$_4$ action is given in terms of the field $J$ and is a sum of two terms, respectively the non-linear $\sigma$-model (NLSM) term  and the Wess-Zumino (WZ) term,
\begin{align}
	S_{WZW_4}[J]=-\frac{i}{4\pi}\int_{\mathbbm{R}^{2,2}}\omega \wedge \text{Tr}[(\partial J)J^{-1}\wedge (\tilde{\partial} J)J^{-1}]-\frac{i}{12\pi}\int_{\mathbbm{R}^{2,2}\times[0,1]}\omega\wedge \text{Tr}[(d\hat{J}\hat{J}^{-1})^3] 
\end{align}
where the two-form is
	$\omega=\frac{i}{2} (dz\wedge d\tilde{z}-dw \wedge d\tilde{w})$, the exterior derivatives are defined as 
$
	d=\partial+\tilde{\partial}+dt \partial_t$, $\partial=dw \partial_w+dz \partial_z$, $\tilde{\partial}=d\tilde{w} \partial_{\tilde{w}}+d\tilde{z} \partial_{\tilde{z}}$
and $\hat{J}(x,t)$ is a homotopy defined such that $\hat{J}(x,1)=J(x)$, $\hat{J}(x,0)=\mathbbm{1}$. 
From the product expression for the multi-pole solution \eqref{eq:psi_prod}, we have a corresponding product form for the $J$ matrix
\begin{align}
	J^{(N)}=\psi^{(N)}(0)=g^{(N)}(0)\dots g^{(1)}(0) ~
\end{align} 
and as the action has the Polyakov-Wiegmann (PW) property \cite{Polyakov:1983tt, Losev:1995cr}
we have that, and here writing $g^{(k)}(0)=g^{(k)}$,
\begin{align}
S_{WZW_4}[J^{(N)}]
&=S_{WZW_4}[J^{(N-1)}]+S_{WZW_4}[g^{(N)}]\nn\\
&\kern+40pt+\frac{i}{2\pi}\int_{\mathbbm{R}^{2,2}} \omega\wedge \text{Tr}[(g^{(N)})^{-1}\tilde{\partial} g^{(N)}
\wedge {\partial} J^{(N-1)} (J^{(N-1)})^{-1}]~.
\end{align}
This can be used recursively to find an expression for the action as a sum of terms each of which is the WZW$_4$ action evaluated on the individual product factors and PW terms involving the coupling to the corresponding background gauge potential. The identification in fact holds pointwise at the level of the Lagrangian densities up to an ambiguity in the WZ term. We will take as our definition 
\begin{align}
	\mathcal{L}_{WZW_4}[J^{(N)}]=\sum_{k=1}^N \left(\mathcal{L}_{WZW_4}[g^{(k)}]+\frac{i\omega }{2\pi} \wedge \text{Tr}[(g^{(k)})^{-1}\tilde{\partial} g^{(k)}
	\wedge {\partial} J^{(k-1)} (J^{(k-1)})^{-1}]\right)
\end{align}
which is natural from the dressing perspective but differs from the local form used in \cite{Hamanaka:2022ole, Li:2025dlc} by an exact term involving the choice of one-form potential for $\omega$. It is worth noting that this definition depends on the ordering of the factorisation through its dependence on the dressing vectors. 
We can simplify this further by using projector identities so that,
\begin{align}
\text{Tr}	[(\partial_a g^{(k)})(g^{(k)})^{-1} (\partial_b g^{(k)})(g^{(k)})^{-1}] = \frac{\kappa_k^2}{|\mu_k|^2}\text{Tr}[\partial_a P_{\xi^k} \partial_b P_{\xi^k}]	
\end{align}
and so we find that the NLSM Lagrangian density for $g^{(k)}$ is proportional to the non-covariant version of $\sigma_k$ \eqref{eq:sigmak}. 
 The PW term provides the correct gauge terms to covariantise the derivatives through
\begin{align}
	(\partial J^{(k)})(J^{(k)})^{-1}=-A^{(k)}_z dz-A^{(k)}_w dw~, ~~~\text{and}~~~A^{(k)}_{\tilde{w}}=A^{(k)}_{\tilde{z}}=0~,
\end{align}
up to an extra term proportional to Re$[\mu_k]$. The WZ terms for $g^{(k)}$ vanish for every $k$ with an appropriate choice of WZ representative. This can be seen by noting that in flat space-time, as $\omega$ is closed, we can choose a one-form $\gamma$, with $d\gamma=\omega$, so that 
\begin{align}
	S_{WZ}[g^{(k)}]=-\frac{i}{12\pi}\int_{\mathbbm{R}^{2,2}} \gamma \wedge \text{Tr}[(dg^{(k)} (g^{(k)})^{-1})^3]~.
\end{align} 
However, as
\begin{align}
P_{\xi^k}dg^{(k)} (g^{(k)})^{-1}P_{\xi^k}=(1-P_{\xi^k})dg^{(k)} (g^{(k)})^{-1}(1-P_{\xi^k})=0~,
\end{align}
 $dg^{(k)} (g^{(k)})^{-1}$ is purely off-diagonal with respect to the splitting Im$P_{\xi^k} \oplus$Ker$P_{\xi^k}$ and so the trace of a triple product must vanish. 
Thus we have that
\begin{align}
	&\mathcal{L}_{WZW_4}[g^{(k)}]+\frac{ i\omega}{2\pi}\wedge \text{Tr}[(g^{(k)})^{-1}\tilde{\partial} g^{(k)}
	\wedge {\partial} J^{(k-1)} (J^{(k-1)})^{-1}]\nn\\
	&\qquad=\frac{-1}{8\pi}
	\left(\frac{\kappa^2_k}{|\mu_k|^2} \sigma_k+
	\frac{ 2 \kappa_k \text{Re}[\mu_k]}{|\mu_k|^2 } 
	\text{Tr}[\partial_{\tilde{z}} P_{\xi^k}A^{(k-1)}_{z}
	-\partial_{\tilde{w}} P_{\xi^k}A^{(k-1)}_{w}]\right)
	\nu
\end{align}
and if we make the choice that all the spectral parameters lie on the positive imaginary axis, so that $\kappa_k=2 i |\mu_k|$, and dropping the overall volume factor $\nu$, we have that
\begin{align}
	\mathcal{L}_{WZW_4}[J^{(N)}]=\frac{1}{2\pi}\sum_k \sigma_k=\frac{1}{2\pi} \Box \log \tau^{(N)}~.
\end{align}
Thus we see that, at least for a simple class of parameter choices, the $\tau$-function acts as a potential for the scalar density defined by the WZW$_4$ action. 

\subsection{Yang-Mills Lagrangian}
We have seen in the one-pole sector that the density $\sigma$ acts as a potential for the Yang-Mills Lagrangian and we checked numerically that this continued to hold for the two-pole sector for a number of particular cases. The result can in fact be shown to hold more generally in the $N$-pole sector by an inductive argument. Specifically, given a $k$-pole gauge field $A^{(k)}$ generated from $A^{(k-1)}$  by the dressing procedure with dressing vector $\xi^k$, the difference of the densities is given by the $k$-th term in $\sigma$:
\begin{align}
\Delta\mathcal{L}=	\mathcal{L}(A^{(k)})-\mathcal{L}(A^{(k-1)})=-\Box \sigma_{k}~.
\end{align}
To show this we recall how the background gauge field changes under the dressing procedure. Substituting $\psi^{(k)}(\zeta)=g^{(k)}(\zeta)\psi^{(k-1)}(\zeta)$ in the Lax equations \eqref{eq:IS_pot}
and expanding in large $\zeta$ we have that
\begin{align}
	A^{(k)}_w=\kappa \partial_{\tilde{z}}P+A_w~, ~~~A^{(k)}_z=\kappa\partial_{\tilde{w}}P+A_z~,~~~A^{(k)}_{\tilde{z}}=A^{(k)}_{\tilde{w}}=0
\end{align}
where for convenience we drop the labels on $P_{\xi^k}$, $\kappa_k$, and $A^{(k-1)}$. In terms of the $K$-matrix, \eqref{eq:Keq}, this corresponds to a shift
\begin{align}
	K^{(k)}=K+\kappa P~.
\end{align}
To find the corresponding change in the Lagrangian density we use the formulation as a sum of total derivatives of potentials which depend on the $K$-matrix \eqref{eq:YML_td} such that the change is
\begin{align}
	\Delta \mathcal{L}=\partial_{\tilde w} \Delta \mathcal{J}_{\tilde{z}}-\partial_{\tilde z} \Delta \mathcal{J}_{\tilde{w}}
\end{align}
with 
\begin{align}
\Delta \mathcal{J}_{\tilde{a}}&=\frac{\kappa}{2} \text{Tr}\big[A_w \partial_{\tilde a} \partial_{\tilde{w}}P+\partial_{\tilde{z}}P\partial_{\tilde{a}}  A_z\big]+\frac{\kappa^2}{2} \text{Tr}\big[\partial_{\tilde{z}}P\partial_{\tilde{a}}  \partial_{\tilde w} P\big]-(\tilde{w}, w\leftrightarrow \tilde{z},z)~.
\end{align}
Thus we wish to show that
\begin{align}
	\partial_{w} \sigma_k=\Delta \mathcal{J}_{\tilde{z}}+\partial_{\tilde{z}} \Lambda~,~~~\partial_{z} \sigma_k=\Delta \mathcal{J}_{\tilde{w}}+\partial_{\tilde{w}} \Lambda~,
\end{align}
for some potential $\Lambda$. This is indeed the case, see App. \ref{app:Lpot} for details, with
\begin{align}
	\Lambda=\frac{\kappa}{2} \text{Tr}(\partial_{\tilde{z}}P A_z-\partial_{\tilde{w}}P A_w)+\text{Re}\,[\mu] \sigma_k~.
\end{align}
Thus we see that for each dressing factor depending on a vector $\xi$ in the product defining $\psi$, we add an additional term $-\Box \sigma_k$ to the Lagrangian density or 
\begin{align}
	\mathcal{L}=-\sum_{k=1}^N \Box \sigma_{k}=-\Box^2 \log \tau~.
\end{align}
The individual terms in the sum depend on the ordering implicit in the dressing construction but the final answer is independent of this choice. 
\section{Conclusions and outlook}

We have seen that for simple-pole solutions, with distinct non-real poles and no conjugate coincidences, on the set where $u^k(x)\neq 0$, we have
\begin{align}
\tau=\prod_k(\bar\mu_k-\mu_k)\det\Gamma>0,
\qquad
\mathcal{L}=-\Box^2\log\tau
\end{align}
and so we interpret this as a $\tau$-function for this class of split-signature solutions. However, this definition is quite tied to the details of the dressing method and it is natural to ask whether there exists something more general. In particular, one may ask if it has a complexified, twistorial interpretation which also produces Osborn's formula in the Euclidean case. Previous work on twistorial constructions \cite{Mason:1994rs, mason2000tau, Mason:2001vj}  identified the logarithmic covariant derivative of a section of a particular determinant line bundle with the logarithmic derivative of the $\tau$-function. It may be useful to study whether the $\tau$ defined here can be identified with this more general construction.  

A motivation for this work is the possibility of a combinatorial description, analogous to that of Kodama and Williams \cite{kodama2011kp}, for interacting wall solitons in ASDYM. While we cannot yet point to any such structure, it is perhaps worth noting that the tropical skeleton described in Sec. \ref{sec:NPS} has a dual description generalising the  triangular Newton polygon of the Y-soliton. Starting from the real wave vectors $k_i=\text{Re}\, p_i$ and the colour-vector norms $h_i=\log \|q_i\|$, we define $n$ points in $\mathbbm{R}^{2,2}\times \mathbbm{R}$. We then take the upper convex hull of these points and project the faces of the hull back to $\mathbbm{R}^{2,2}$ which defines the regular subdivision $\mathcal{S}_h$ of the point configuration $\{k_i\}$ induced by the heights $h_i$. Then
\begin{eqnarray}
	&\text{regions } R_i \ \leftrightarrow\ \text{vertices of } \mathcal{S}_h~,\qquad &
	\text{walls } [ij] \ \leftrightarrow\ \text{edges of } \mathcal{S}_h~,\qquad\nn\\	
	&\text{two-surfaces } [ijk] \ \leftrightarrow\ \text{2-faces of } \mathcal{S}_h~,\nn\qquad &
	\text{lines } [ijkl] \ \leftrightarrow\ \text{3-faces of } \mathcal{S}_h~,\\
		& &\kern-100pt\text{vertices } [ijklm] \ \leftrightarrow\ \text{4-cells of } \mathcal{S}_h~  .
\end{eqnarray}
It would be interesting to understand if this, or some other geometrical structure, can provide insight into the one-pole or ideally all multi-pole solutions. One notable feature which differs from the geometry of KP-II solutions is that of positivity. In KP-II positivity is required for regular solutions while here, except for singular points, the $\tau$-function is guaranteed to be positive. It may be necessary to better understand the notion of positivity for ASDYM solutions to make progress in this direction.

While the wall solitons are generically interacting with non-trivial phase shift, the lump solitons only show scattering in restricted kinematics. Specifically, starting from the two-pole solution with momentum spinors $\tilde{\lambda}_i=(1, \mu_i)$ the double-pole limit is $\mu_1, \mu_2\to \mu$  which implies that $[\tilde{\lambda}_1, \tilde{\lambda}_2] \to 0$.  This is the half-collinear limit used in the construction of single-minus tree-level scattering amplitudes  \cite{Guevara:2026qzd}. The scattering lump solution further sets the unprimed spinors equal, $\lambda_1=\lambda_2=(a_\ell, a_m)$, and so it is in fact in the completely collinear limit. However, we can construct a solution with half-collinear but distinct momenta by allowing separate holomorphic dependence on both $\ell$ and $m$, taking 
\begin{align}
	f=f_0+c_1 e^{\theta_1}~, ~~~h=h_0+c_2 e^{\theta_2}~,~~~\theta_i=p_i\cdot x
\end{align}
and $g\to H=(w\partial_\ell+z\partial_m)f+h$. 
This is no longer in the class of solutions we have previously considered but one can check that $K=\kappa(P_1+P_2)$, using \eqref{dp_proj}, satisfies the equation of motion \eqref{eq:Keq}. 
It may be possible to generalise this to solutions depending on more momenta and use them to compute higher-point scattering amplitudes via perturbiner methods \cite{Rosly:1996cp, Rosly:1996vr}.

\paragraph{Acknowledgements}
I would like to thank the participants of the workshops \textit{Higher-d Integrability},
Favignana, 2025 and \textit{Higher-dimensional integrability and holography}, Buggerru, 2026 for fruitful discussions on topics related to this work. I thank Shangshuai Li, Ken-ichi Maruno and Da-jun Zhang for sharing their preprint \cite{li202621dim} and Masashi Hamanaka for interesting comments. I would particularly like to acknowledge discussions with Lewis Cole which contributed significantly to the approach taken in this work. I also acknowledge the use of large language models in the development and preparation of this work, with access provided by the OpenAI Academic Researchers programme and Claude Team plan for Scientists; however, I take full responsibility for the accuracy of this work, having reviewed, verified, and approved all AI-generated content. Finally, I would like to thank the CERN Theory Department for their hospitality while this work was completed. 
\appendix

\section{Notation and conventions}
\label{app:notations}
It is convenient to consider the theory defined on complex Minkowski space $\mathbbm{C}M$ with double-null coordinates in which the metric is 
\begin{align}
	ds^2=2dzd\tilde{z}-2dw d\tilde{w}
\end{align}
and the volume element is $\nu=dw\wedge d\tilde{w} \wedge dz \wedge d\tilde{z}$. To recover real spaces we must impose reality conditions on the coordinates.
There are two methods to find the ultrahyperbolic slice or Klein space $\mathbbm{R}^{2,2}$. One is to impose $\bar{z}=\tilde{z}$ and $\bar{w}=\tilde{w}$ which gives the signature $(++--)$. A second with signature $(+--+)$, which we will use, is to take all coordinates to be real. In this case we can relate the coordinates as 
\begin{align}
	\begin{pmatrix}\tilde{z} & w \\ \tilde{w} & z \end{pmatrix}=\frac{1}{\sqrt{2}} \begin{pmatrix} x^0+ x^1 & x^2- x^3 \\ x^2+ x^3 & x^0- x^1 \end{pmatrix}. 
\end{align}
The volume element is given by 
\begin{align}
	d^4x=dx^0\wedge dx^1 \wedge dx^2 \wedge dx^3
\end{align}
which is real. We define $p\cdot x=p_\mu x^\mu=g^{\mu\nu} p_\mu x_\nu$ as standard but use a non-standard normalisation for $\Box=\tfrac{1}{2} g^{\mu\nu}\partial_\mu\partial_\nu$.

It will also be convenient in places to use spinor notation where we identify the space of four-vectors with the tensor product of left- and right-handed spinors $S$ and $S'$. Concretely we write the vector $x^\mu=(w, z, \tilde{w}, \tilde{z})$ as a two-index spinor
\begin{align}
	x^{AA'}=\begin{pmatrix} x^{00'} & x^{01'} \\ x^{10'} & x^{11'} \end{pmatrix}=\begin{pmatrix}\tilde{z} & w \\ \tilde{w} & z \end{pmatrix}~.
\end{align}
We can correspondingly identify covectors with tensor products of dual space spinors
and in particular coordinate derivatives $\partial_\mu$ become spinor operators
\begin{align}
	\partial_{AA'}=\begin{pmatrix} \partial_{00'} & \partial_{01'} \\ \partial_{10'} & \partial_{11'} \end{pmatrix}=\begin{pmatrix}\partial_{\tilde{z}} & \partial_w \\ \partial_{\tilde{w}} & \partial_z \end{pmatrix}~.
\end{align}
We can introduce skew-symmetric two-index spinors $\epsilon_{AB}$, $\epsilon_{A'B'}$, $\epsilon^{AB}$, $\epsilon^{A'B'}$, with $\epsilon_{01}=\epsilon_{0'1'}=\epsilon^{01}=\epsilon^{0'1'}=+1$, which can be used to raise and lower spinor indices,  
\begin{align}
	\alpha_B=\alpha^A \epsilon_{AB}~,~~~ \beta^A=\epsilon^{AB} \beta_B
\end{align}
and similarly for primed indices. 

For the polarisation vectors we have the standard Hermitian inner product and norm
\begin{align}
	\lpangle q_1, q_2\rpangle =q_1^\dagger q_2=\bar{q}_{1a}q_{2a}~, ~~~\|q\|^2=(q,q)~.
\end{align}

\section{Determinant factorisation}
\label{app:detfact}
Consider the $N$-soliton matrix
\begin{align}
	\Gamma_N=\begin{pmatrix} \Gamma_{N-1} & \delta\\ 
		-\delta^\dagger & \Gamma^{NN}\end{pmatrix}
\end{align}
where the corner entry $\Gamma^{NN}=\|u^N\|^2/(\bar{\mu}_N-\mu_N)$ and the additional column vector
\begin{align}
	\delta^k=\frac{\lpangle u^k,u^N\rpangle}{\bar{\mu}_k-\mu_N}~
\end{align}
is related to the additional row vector by anti-Hermiticity. 
The determinant of $\Gamma_N$ factorises
\begin{align}
	\det \Gamma_N=\det \Gamma_{N-1} \cdot s_N
\end{align}
where $s_N=\Gamma^{NN}+\delta^\dagger \Gamma_{N-1}^{-1} \delta$ is the Schur complement of $\Gamma_{N-1}$ in $\Gamma_{N}$. Now we can show that $s_N$ is given by the norm of the $N$-th dressing vector $\xi^N$:
\begin{align}
	\xi^N&=\psi^{(N-1)}(\bar{\mu}_N)\bar{u}^N
	=\bar{u}^N+\sum_{k,k'=1}^{N-1} (\delta^\dagger)^{k}(\Gamma_{N-1}^{-1})^{kk'} \bar{u}^{k'}~.
\end{align}
The norm of $\xi^N$ is 
\begin{align}
	\|\xi^N\|^2&=\|u^N\|^2-\delta^\dagger(D-\mathbbm{1}\bar{\mu}_N)\Gamma_{N-1}^{-1} \delta+\delta^\dagger \Gamma_{N-1}^{-1}(\bar{D}-\mathbbm{1}{\mu}_N)\delta -\delta^\dagger\Gamma_{N-1}^{-1} (\bar{D}\Gamma_{N-1}-\Gamma_{N-1}D)\Gamma_{N-1}^{-1}\delta\nn\\
	&=-\kappa_N(\Gamma^{NN}+\delta^\dagger \Gamma_{N-1}^{-1}\delta)
\end{align}
where we have used the notation $D=\text{diag}(\mu_1, \dots, \mu_{N-1})$. Hence we have that
\begin{align}
	\det \Gamma_N=(\|\xi^N\|^2/\bar{\kappa}_N )\det \Gamma_{N-1}~.
\end{align}

\section{Computation of $\Lambda$}
\label{app:Lpot}
In order to determine $\Lambda$ we proceed by direct computation. To this end, we derive an alternative expression for $\sigma_\xi$ in terms of derivatives of the projection operator, $P=\frac{\xi\otimes \xi^\dagger}{\|\xi\|^2}$, with respect to the directions with trivial gauge potential, namely $\tilde{z}$ and $\tilde{w}$, and then compute second derivatives. 
The projector $P$ for any vector satisfies, 
\begin{align}
	P_a=\partial_a\left(\frac{\xi\otimes \xi^\dagger}{\|\xi\|^2}\right)=\frac{(1-P)\xi_a\otimes \xi^\dagger+\xi\otimes \xi_a^\dagger(1-P)}{\|\xi\|^2}
\end{align}
where we use the notation $X_a=\partial_a X$, and hence we have that
\begin{align}
	\label{eq:proj1}
	P P_a=\frac{\xi\otimes \xi_a^\dagger(1-P)}{\|\xi\|^2}~,~~~ P_a P=\frac{(1-P)\xi_a\otimes \xi^\dagger}{\|\xi\|^2}~.
\end{align}
This implies
\begin{align}
	PP_aP=(1-P) P_a (1-P)=0
\end{align}
so that $P_a$ is off-diagonal with respect to the decomposition Im$P\oplus$Ker$P$ which is useful in simplifying certain traces of products of projection operators and their derivatives. A similarly useful identity is 
\begin{align}
	\text{Tr}[P[[P,Y],P_{a}]]=\text{Tr}[YP_a]
\end{align}
for any $Y$ as $[P,[P,Y]]$ projects $Y$ onto its off-diagonal components. 
For dressing vectors $\xi$, which satisfy
the conditions
\begin{align}
	D_a\xi =\bar{\mu} \partial_{\hat{a}}\xi~,~~~a=(z,w)
\end{align}
where we have introduced the further notation 
\begin{align}
	\partial_{\hat{a}}=\delta_{a}^{\tilde a}\partial_{\tilde{a}}~,~~~\text{with}~~~\delta_{w}^{\tilde{w}}=\delta_{z}^{\tilde{z}}=0~,~~~ \delta_{w}^{\tilde{z}}=\delta_{z}^{\tilde{w}}=1~,
\end{align}
the relations \eqref{eq:proj1} can be used to show that
\begin{align}
	\sigma_\xi=\Box \log\|\xi\|^2=-\kappa\text{Tr}(P[P_{\tilde{z}},P_{\tilde{w}}])~.
\end{align}
We can similarly compute the covariant derivatives of projection operators of dressing vectors in terms of derivatives in gauge trivial directions
\begin{align}
	D_a P=\frac{\kappa}{2}[P, P_{\hat{a}}]+\text{Re}[\mu]\,\partial_{\hat{a}} P~,
\end{align}
so that second derivatives are given by
\begin{align}
	\nabla_a P_{\tilde {b}} \equiv 
	(D_a-\text{Re}(\mu) \partial_{\hat{ a}} )\partial_{\tilde{b}}P
=\frac{\kappa}{2}
\left(
[P_{\tilde{b}}, P_{\hat{a}}] +[P,P_{\tilde{b} \hat{a}}]	\right)+[P, \partial_{\tilde{b}}A_a]
\end{align}
where $\tilde{b}=(\tilde{z},\tilde{w})$.
A slightly longer calculation shows that
\begin{align}
\nabla_a \sigma_\xi\equiv (\partial_a-\text{Re}[\mu]\partial_{\hat{a}})\sigma_\xi=-\kappa \text{Tr}[Y_{\tilde{z}a} P_{\tilde{w}}-Y_{\tilde{w}a} P_{\tilde{z}}]
\end{align}
with 
\begin{align}
Y_{\tilde{b}a}=\frac{\kappa}{2} P_{\tilde{b} \hat{a}}+\partial_{\tilde{b}}A_a~.
\end{align}
Finally one has that
\begin{align}
	\nabla_a \sigma_\xi=\Delta \mathcal{J}_{\hat a}+\frac{\kappa}{2}\partial_{\hat{a}}T
\end{align}
where 
\begin{align}
	T=\text{Tr}(P_{\tilde{z}}A_{z}-P_{\tilde{w}}A_w)~.
\end{align}

\bibliographystyle{utphys}

\bibliography{SDYM.bib}

\end{document}